\documentclass[10pt, journal, final]{IEEEtran}

\usepackage{amsfonts}
\usepackage{graphicx}
\usepackage{subfigure}

\usepackage{amsmath}
\usepackage{amssymb}
\usepackage{epsf}
\usepackage{alltt}

\usepackage{multirow}
\usepackage{cite}
\usepackage{listings}
\usepackage{flushend}

\usepackage{xcolor}
\definecolor{dkgreen}{rgb}{0,0.6,0}
\definecolor{gray}{rgb}{0.5,0.5,0.5}
\definecolor{mauve}{rgb}{0.58,0,0.82}
  
\newtheorem{cor}{Corollary}
\newtheorem{lem}{Lemma}
\newtheorem{definition}{Definition}
\newtheorem{thm}{Theorem}

\usepackage{xcolor}

\newcommand{\eq}[1]{Eq.~(\ref{#1})} 

\begin{document}

\thispagestyle{empty}
\bibliographystyle{siam}

\title{Adiabatic Quantum Programming:\\Minor Embedding With Hard Faults}

\author {
Christine~Klymko*,\thanks{*Department of Mathematics and Computer Science,
Emory University, Atlanta, Georgia 30322, USA
(cklymko@emory.edu).}
Blair~D.~Sullivan\textsuperscript{\textdagger}\thanks{\textsuperscript{\textdagger}Computer Science and Mathematics Division, Oak Ridge National Laboratory, 1 Bethel Valley Road, Oak Ridge, Tennessee 37831, USA 
(sullivanb@ornl.gov, humblets@ornl.gov).},
and~Travis~S.~Humble\textsuperscript{\textdagger}
}

\maketitle

\markboth{{\sc C.~Klymko, B.~D.~Sullivan, and T.~S.~Humble}}{{Minor Embedding With Hard Faults}}

\begin{abstract}
Adiabatic quantum programming defines the time-dependent mapping of a quantum algorithm into an underlying hardware or logical fabric. An essential step is embedding problem-specific information into the quantum logical fabric. We present algorithms for embedding arbitrary instances of the adiabatic quantum optimization algorithm into a square lattice of specialized unit cells. These methods extend with fabric growth while scaling linearly in time and quadratically in footprint. We also provide methods for handling hard faults in the logical fabric without invoking approximations to the original problem, and illustrate their versatility through numerical studies of embeddabilty versus fault rates in square lattices of complete bipartite unit cells. The studies show these algorithms are more resilient to faulty fabrics than naive embedding approaches, a feature which should prove useful in benchmarking the adiabatic quantum optimization algorithm on existing faulty hardware.
\end{abstract}

\begin{keywords} 
quantum computing, adiabatic quantum optimization, graph embedding, fault-tolerant computing
\end{keywords}

\section{Introduction}
Adiabatic quantum optimization (AQO) applies the principles of quantum computing to solve unconstrained optimization problems. In particular, the AQO algorithm interpolates between two quantum logical Hamiltonians in order to adiabatically transform an initial quantum state to a computational solution state \cite{Farhi2001}. This specialized application of adiabatic quantum computing has been used to solve a variety of problems including, for example, instances of satisfiability (SAT) \cite{Farhi2000} and exact cover \cite{Farhi2001}, finding Ramsey numbers \cite{Gaitan2011}, classifying binary images \cite{Neven2008}, training classifiers for machine learning \cite{Pudenz2012} and finding the lowest free-energy configuration in folded proteins \cite{PerdomoOrtiz2012}. 
\par
Benchmarking the efficiency of the AQO algorithm is currently of significant interest in quantum computer science. Whereas some studies of optimization problems have uncovered runtimes that scale polynomially in problem size, others suggest worst-case exponential behavior, or even trapping in local minima \cite{Altshuler2010}. Interpreting these analyses are difficult, in part, because of the manner in which instance-specific information alters the implementation of the algorithm, i.e., programming. As emphasized by others \cite{Farhi2000,Dickson2011,Altshuler2010,Dickson2012}, choices made in programming the AQO algorithm greatly impact its runtime and, consequently, the observed scaling behavior. 
\par
Benchmarking adiabatic algorithms is further complicated when the design of the logical Hamiltonians is constrained. Because the AQO algorithm uses a reduction of the classical optimization problem to a quantum logical representation, i.e., a Hamiltonian, any constraints placed on this underlying logical fabric can only limit performance. Understanding the impact of the logical fabric is especially pertinent since existing AQO hardware supports a specific topology over a relatively modest number of qubits \cite{8qubit, DWave}.
\par
Adiabatic quantum programming has been described previously as requiring two steps: parameter setting \cite{Choi1} and minor embedding \cite{Choi2}. Minor embedding, in particular, uses explicit information about the logical fabric as well as the problem to generate the implementation of the AQO algorithm. Choi has demonstrated how an arbitrary input graph can be minor embedded within one type of highly regular fabric, a square lattice of $K_{4,4}$'s, complete bipartite graphs with eight vertices. In the current paper, we also present algorithms for minor embedding into additional logical fabrics, namely, square lattices of $K_{c,c}$ with $c \geq 1$.  We present an attempt at a brute force embedding via graph isomorphism in maximal minors  (henceforth called maximal minor embedding) as well as an algorithm for complete-graph embedding. We compare these algorithms in terms of their complexity as well as the scaling of the embedding result.
\par
Notwithstanding algorithms for the unit-cell lattice, an open question in adiabatic quantum programming is how to handle fabrics containing randomized hard faults. Hard faults refer to defects in the logical fabric that compromise its regularity. As their locations are random, the embedding algorithm must handle a variety of target graphs. In the current paper, we present methods for minor embedding that use heuristics to adapt to random faults in the logical fabric (hardware). We analyze algorithmic performance in terms of the maximum embeddable complete graph obtained using numerical simulations. These studies quantify the impact of faults on the required logical footprint and provide performance expectations for hard fault-tolerant adiabatic quantum programming.
\par
The paper is organized as follow: Sec.~II defines the role of minor embedding in adiabatic quantum optimization; Sec.~III briefly reviews previous work; Sec~IV defines nomenclature and presents implications of treewidth on graph embeddability; Sec.~V recounts properties of the unit-cell lattice; Sec.~VI determines treewidth for hardware graphs $F$; Sec.~VII presents embedding of a complete graph in $F$; Sec.~VIII presents two algorithms for embedding with hard faults and numerical tests of these algorithms using randomized fault placement; finally, Sec.~IX presents our conclusions.

\section{Adiabatic Quantum Optimization}
The AQO algorithm is based on the reduction of an unconstrained optimization problem to a quantum logical Hamiltonian that is diagonal in the computational basis \cite{Farhi2001}. The reduction most naturally begins in terms of binary variables that can then be mapped to the qubits of a logical Hamiltonian $H_F$. For AQO, the problem Hamiltonian takes the form
\begin{equation}
\label{HF}
H_F = \sum_{i \in V_F}{\alpha_{i} Z_{i}} + \sum_{(i,j) \in E_F}{\beta_{i,j} Z_{i} Z_{j}},
\end{equation}
where $\alpha_i$ is the weight on the $i$-th qubit, $\beta_{i,j}$ is the coupling between qubits $i$ and $j$, and the sets $V_F$ and $E_F$ denote the vertices and edges of the  graph $F$ describing the logical fabric; a more formal definition of the hardware graph is found in Sec. \ref{sec:BGT}. In this setting, the Pauli $Z_i$ operator defines the computational basis for the $i$-th qubit.
\par
The 2-local form of \eq{HF} restricts the optimization problems that can be mapped directly into $H_F$. Specifically, any binary optimization problem can be recast to have at most quadratic interactions, i.e., as a quadratic unconstrained binary optimization (QUBO) problem. This reduction can be done by, e.g., substituting the product of two variables with a new one and adding a corresponding penalty term \cite{Boros2002}. The AQO program input is therefore defined as the QUBO problem
\begin{equation}
\label{QUBO}
\arg\min_{{\bf x}\in {\bf B}^{n}}{{\bf x^{T}}{\bf P} {\bf x}},
\end{equation}
where \textbf{x} is a vector of $n$ binary variables and \textbf{P} is an $n$-by-$n$ symmetric real-valued matrix. 
\par
In programming the QUBO problem, the interactions between variables represented by \textbf{P} must be mapped into the quantum logical fabric. We interpret \textbf{P} as a weighted version of the adjacency matrix of an input (problem) graph $P$ describing these dependencies. Hence, programming the AQO algorithm requires embedding $P$ in the graph $F$ representing the logical fabric. We defer the formal definition of minor embedding to Sec.~II.B, but it suffices to say that this yields a graph $F^* = (V^*, E^*)$ contained within the logical fabric, over which a Hamiltonian $H_{F^*}$ is defined as
\begin{equation}
H_{F^*} = \sum_{i \in V^*}{\alpha^*_{i} Z_{i}} + \sum_{(i,j) \in E^*}{\beta^*_{i,j} Z_{i} Z_{j}}
\end{equation}
with $\alpha^*_i$ and $\beta^*_{i,j}$ the corresponding weights and couplings. Setting these parameters requires both the matrix $\textbf{P}$ and the embedding into the logical fabric specified by $F^*$ \cite{Choi1}.
\par
The program for the AQO algorithm is then expressed by the time-dependent Hamiltonian
\begin{equation}
H(t;T) = A(t;T)H_{I} + B(t;T)H_{F^*},
\end{equation}
where $A(t)$ and $B(t)$ control the time-dependent interpolation between an initial Hamiltonian $H_{I}$ and the final embedded problem Hamiltonian $H_{F^*}$. The time $T$ represents the annealing time of the algorithm, such that $H(T) = H_{F^*}$. Running the program $H(t)$ requires initializing the quantum register state to be a ground state of $H(0)$. This is followed by annealing to the time $T$ after which the register is measured. Provided the conditions of the adiabatic theorem are met, the state of the register at $T$ will be a ground state of $H_{F^*}$ and a solution to the QUBO problem. In order to meet these conditions, $T$ must scale inversely with the minimum spectral gap of $H(t)$ \cite{Farhi2001}. The gap, of course, depends on the programmed implementation and we may expect that the choice of embedding plays a role in satisfying this condition.

\section{Previous Related Work}
In \cite{Choi2}, Choi described a hardware graph for minor embedding a large clique, $K_n$, in a limited number of qubits.  This layout was called TRIAD.  Choi also discussed using the TRIAD scheme on a 128 qubit hardware made up of a $4\times4$ grid of $K_{4,4}$ cells to achieve the embedding of $K_{17}$. We note the figure in \cite{Choi2} corresponding to this description in that paper only embeds a $K_{16}$, but  it is possible to embed $K_{17}$ using the TRIAD scheme. It is also worth noting that while the text claimed a requirement of only 6 physical vertices for each logical qubit, this is not achievable with the given hardware (and is not realized in the example given). Our work results in the same embedding for $K_{17}$ on the $4\times4$ grid, but then extends the algorithm to work on a large family of related logical fabrics. We also provide a straightforward algorithm for extending an embedding from an $n\times n$ grid to an $(n+1)\times (n+1)$ grid of $K_{c,c}$  cells. This paper additionally determines the treewidth of the family of fabric graphs, which enables better screening of QUBOs for feasible embeddability. Perhaps most importantly, prior work did not consider the case of faulty fabric, which we address with two algorithms and a set of simulations to demonstrate performance.

\section{Graph Minors and Tree-decomposition}\label{sec:BGT}
A {\em graph} $G=(V,E)$ is a set of vertices $V$ and a set of edges $E$ formed by unordered pairs of vertices.  In this paper, all graphs are finite, simple (no loops or multiple edges), and undirected.  A graph $H = (W,F)$ is a {\em subgraph} of $G$, denoted $H \subseteq G$, if $W \subseteq V$ and $F \subseteq E$.
\par
A {\em path} in $G =(V,E)$ is a sequence of vertices $v_1, v_2, \ldots, v_k$ such that for $1 \leq i < k$, $(v_i, v_{i+1}) \in E$.  A {\em cycle} is a path where $v_1 = v_k$.  If there are no repeated vertices, the path (cycle) is a {\em simple path (cycle)}.
\par
A graph is {\em connected} if there is a path from $u$ to $v$ for every pair of distinct vertices $u,v$ in $V$. A {\em tree} is a connected graph which does not contain any simple cycles as subgraphs. We say a graph $H$ is a {\em subtree} of $G$ if $H \subseteq G$ and $H$ is a tree.
\par
Programming adiabatic quantum computing hardware to solve a specific problem requires embedding a {\em problem graph} $P = (V_P, E_P)$ representing the QUBO problem (elements of $V_P$ correspond to QUBO variables and $E_P = \{(i,j) | \textbf{ P}_{i,j} \neq 0 \}$) into a {\em hardware graph} $F=(V_F, E_F)$ whose vertices representing the qubits and edges are determined by couplings in the logical fabric. In some cases, this can be done in a one-to-one manner through subgraph embedding. 

\begin{definition}
\label{def:subgraph}
 A {\em subgraph embedding} of $P$ into $F$ is a mapping $f: V_P \rightarrow V_F$ such that:
\begin{itemize}
	\item each vertex in $V_P$ is mapped to a unique vertex in $V_F$.
	\item if $(u,v) \in E_P$, then $(f(u), f(v)) \in E_F$. \\
\end{itemize}
Note that if such an $f$ exists, $P$ is a {\em subgraph} of $F$, $P \subseteq F$.
\end{definition}
However, due to design constraints on the underlying logical fabric, in order to consider a large class of QUBO problems, $P$ will need to be embedded into $F$ as a minor. 

\begin{definition}
\label{def:minor}
A {\em minor embedding} of $P$ in $F$ is defined by a mapping $\phi: V_P \rightarrow V_F$ such that:
\begin{itemize}
	\item each vertex $v$ in $V_P$ is mapped to the vertex set of a connected subtree $T_v$ of $F$.
	\item if $(u,v) \in E_P$, then there exist $i_u, i_v \in V_F$ such that $i_u \in T_u$, $i_v \in T_v$, and $(i_u,i_v) \in E_F$.\\
\end{itemize}
If such a mapping $\phi$ exists, then $P$ is {\em minor-embeddable} in $F$ or $P$ is a {\em minor} of $F$, written $P\leq_m F$.  
\end{definition}

Equivalently, $P$ is minor-embeddable in $F$ if $P$ can be obtained from $F$ by a series of edge deletions and contractions (see \cite{Diestel} for more information on graph minors).  Note that every subgraph embedding is also a minor embedding (since $f(v)$ is a single node subtree of $F$). Furthermore, the property of being a minor is transitive: $G \leq_m F$ and $P \leq_m G$ implies $P \leq_m F$. 

Closely related to the idea of a graph minor is the concept of a tree decomposition, a combinatorial way of measuring how ``tree-like"  a graph is. Many early results on graph minors were first proved for trees \cite{Diestel}.  Additionally, certain problems which have exponential complexity on arbitrary graphs have been shown to have polynomial complexity on graphs of bounded treewidth.  More importantly, certain properties of tree decompositions, including upper bounds on treewidth (the definition of which can be found below), are closed under the taking of minors.  Understanding the tree decomposition of the hardware graph gives us information about the properties of the minors the graph has and, thus, what sort of QUBO problems can be embedded.

\begin{definition}
\label{def:treedecomp} Given a graph $G=(V, E)$ let $T=(I,D)$ be a tree, and $\mathcal{V}=\{V_t\}_{i \in I}$ be a family of vertex sets (also called {\em bags}) with $V_i \subseteq V$ indexed by the elements of $I$.   The pair $(T, \mathcal{V})$ forms a {\em tree decomposition} of $G$ if the following hold:
\begin{enumerate}
	\item $V = \cup_{i \in I} V_i$.
	\item if $(u,v) \in E$, then there exists $i \in I$ such that $\{u, v\} \subseteq V_i$.
	\item  for $i_1, i_2, i_3 \in I$, if $i_3$ lies on the path in $T$ between $i_1$ and $i_2$, then $V_{i_1} \cap V_{i_2} \subseteq V_{i_3}$.  
Equivalently, for any vertex $v \in V$, $\{i:\, v \in V_i\}$ forms a connected subtree of $T$. \\
\end{enumerate}
\end{definition}

To avoid confusion, the elements of $V$ are referred to as the {\em vertices} of G and the elements of $I$ as the {\em nodes} of $T$.  The {\em width} of a tree decomposition $(T, \mathcal{V})$ is given by $\max_{i \in I} \{|V_i| - 1\}$.  The {\em treewidth} $\tau(G)$ of a graph $G$ is the minimum width over all tree decompositions of $G$.  Note that the width of any tree decomposition of $G$ gives an upper bound on $\tau(G)$. The following lemmas are well-known in graph theory and are useful for using treewidth to analyze the quantum hardware graphs described in Sec. \ref{sec:hardwaregraph}. \\
\begin{lem}
\label{lem:twlowerbound}
	If $H$ is a minor of $G$ (i.e. $H$ is minor-embeddable in $G$), then $\tau(H) \leq \tau(G)$. \\
\end{lem}

Thus, given the treewidth of a logical fabric $F$, it is possible to automatically narrow down the class of QUBO problems for which it may be possible to find an embedding. The treewidth of several classic families of graphs is known exactly:\\
\begin{lem} \label{lem:twKn,n}\label{lem:twKn}  Let $K_n$ be the complete graph on $n$ vertices and $K_{n,n}$ the complete bipartite graph on $2n$ vertices.
\begin{enumerate}
\item  $\tau(K_n) = n-1$. 
\item $\tau(K_{n,n}) = n$.
\item The treewidth of an $n\times m$ 2-D planar grid is given by $\min\{m,n\}$. 
\end{enumerate}
\end{lem}

For more information on tree decomposition and graph minors (including the proofs of the above lemmas) see \cite{Bodlaender93}, chapter 12 of \cite{Diestel}, and \cite{Fomin2003}.

\section{Description of Hardware graph}
\label{sec:hardwaregraph}
\begin{figure}[t!]
\centering
\includegraphics[width=0.35\textwidth]{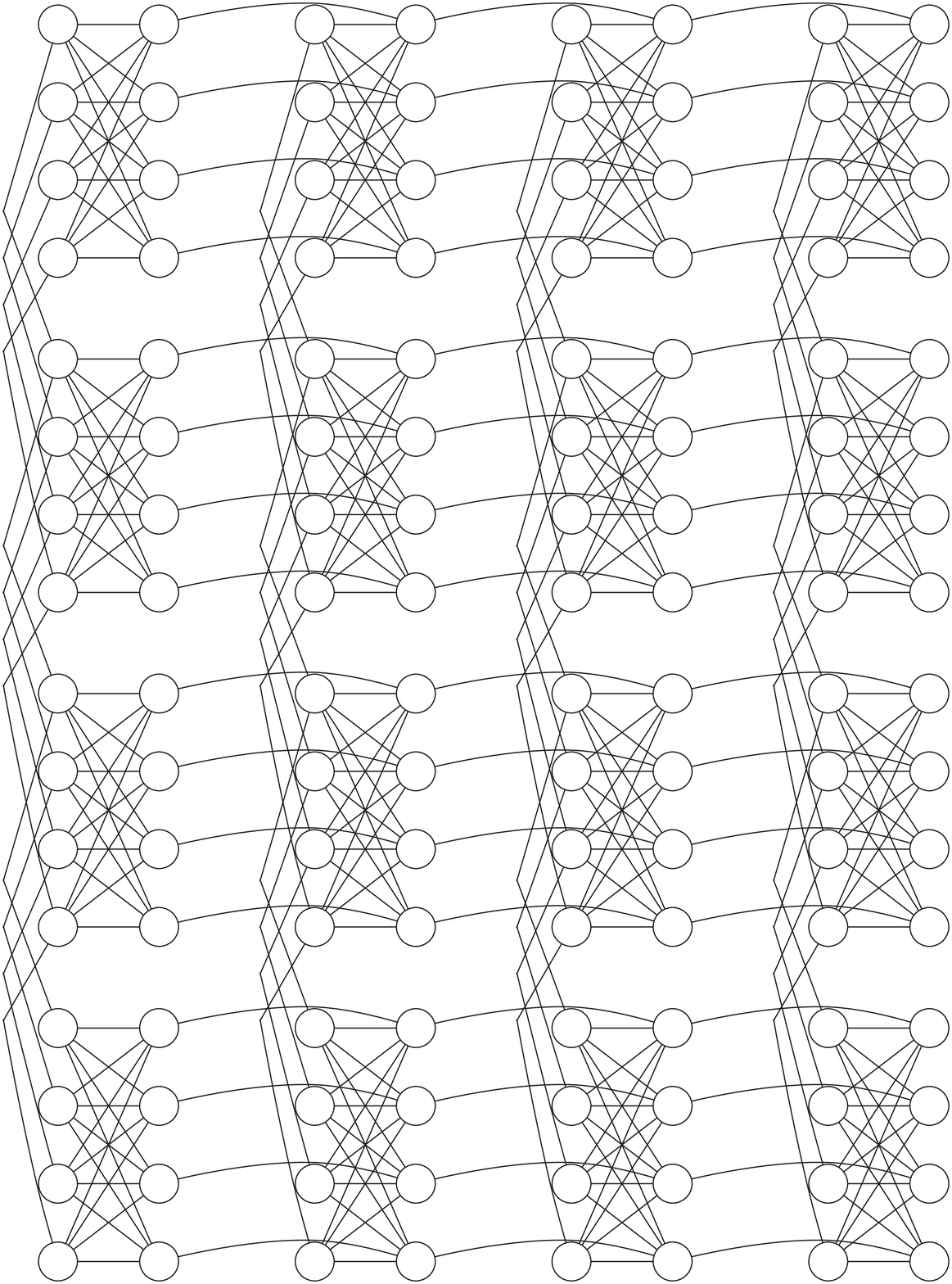}
\caption{A 4$\times$4 array of $K_{4,4}$ unit cells coupled as in the hardware graph from \cite{8qubit}.}
\label{fig:blankhardware}
\end{figure}

In this section, we review the hardware graph that has been the basis for several proposed or demonstrated experimental studies \cite{Gaitan2011,Bian2012,PerdomoOrtiz2012,DWave}. The building blocks of this graph are 8-qubit unit cells whose internal couplings form $K_{4,4}$ \cite{8qubit}. Unit cells are tiled together with each qubit on the left half of a $K_{4,4}$ connected to its image in the cells directly above and below, and each qubit on the right half of the $K_{4,4}$ connected to its image in the cells directly to the left and right.  A representation of the graph formed by sixteen cells is shown in Fig. \ref{fig:blankhardware}. Note that due to the way the qubits are physically connected \cite{8qubit}, when there is a failure, it will be the failure of a qubit and not an individual coupler.  In terms of the hardware graph, this means vertices (and all their adjacent edges) will fail, not individual edges.

In our analysis, we consider extensions of the unit cell design to include an increase in the number of qubits forming a cell. We also parameterize the hardware fabric to allow for expanding the grid of unit cells. In general, our results are applicable in the setting where cells consist of $2c$ qubits forming a $K_{c,c}$ and are attached to form an $m\times m$ grid in the same manner as described above.  We denote a hardware graph of this form as $F(m,c)$. For example, the hardware graph shown in Fig.~\ref{fig:blankhardware} corresponds to $F(4,4)$.
\par
For ease of reference, we define a labeling on $V_{F(m,c)}$. First, we number a single cell: the vertices on the left half of the $K_{c,c}$ as $1, 2, \ldots, c$ from top to bottom, and the vertices on the right half of the $K_{c,c}$ as $c+1, c+2, \ldots, 2c$, again from top to bottom.  See Fig. \ref{subfig:qubitnumbering} for an example of this numbering in a $K_{4,4}$ cell. Each vertex in $V_{F(m,c)}$ is then given a label of the form $v_{a,b}^d$ where $(a,b)$ is the (row, column) position of the cell containing the vertex in the $m\times m$ 
grid - with cell (1,1) in the upper left corner - and $d$ corresponds to the position of the vertex inside the individual cell, as described above.

\section{Treewidth of the Hardware graph}
As seen in Lemma \ref{lem:twlowerbound}, if the treewidth of the hardware graph is known, it can be used to a priori rule out the possibility of embedding certain classes of QUBOs.

\begin{figure*}[th!]
\centering
\subfigure[]{
	\includegraphics[width=.29\textwidth]{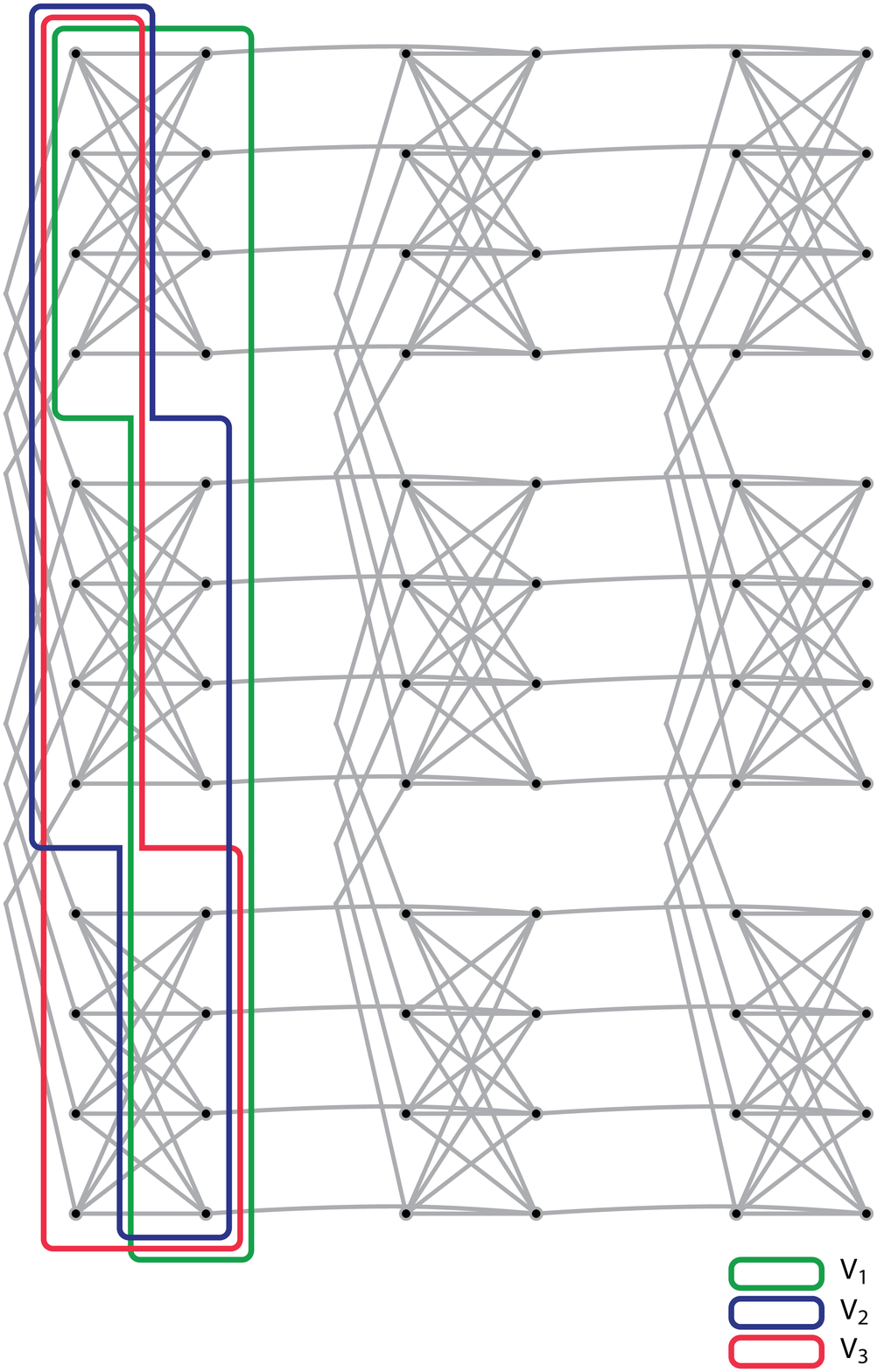}
}
\hspace{0.3in}
\subfigure[]{
	\includegraphics[width=0.29\textwidth]{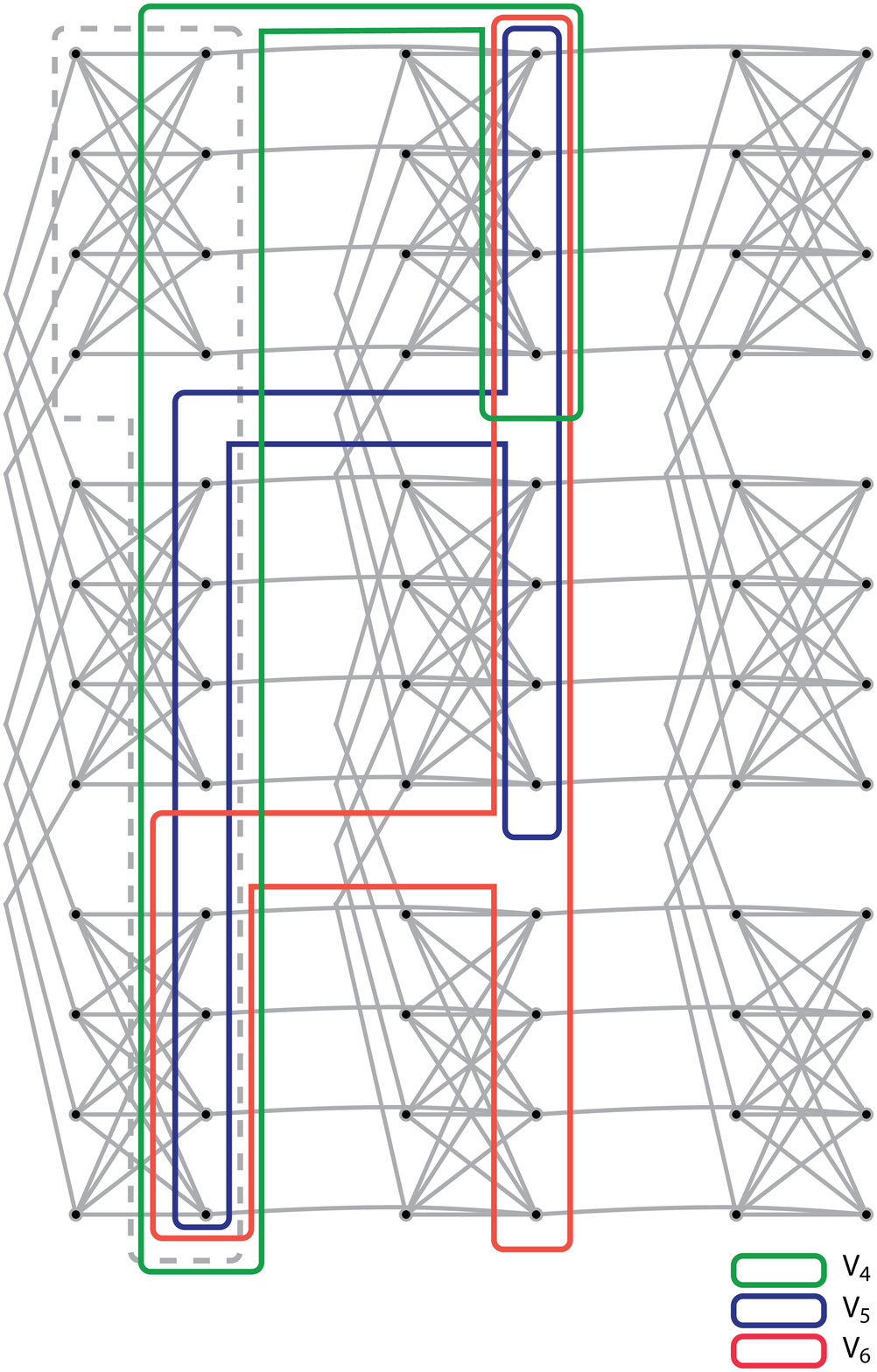}
}

\subfigure[]{
	\includegraphics[width=0.29\textwidth]{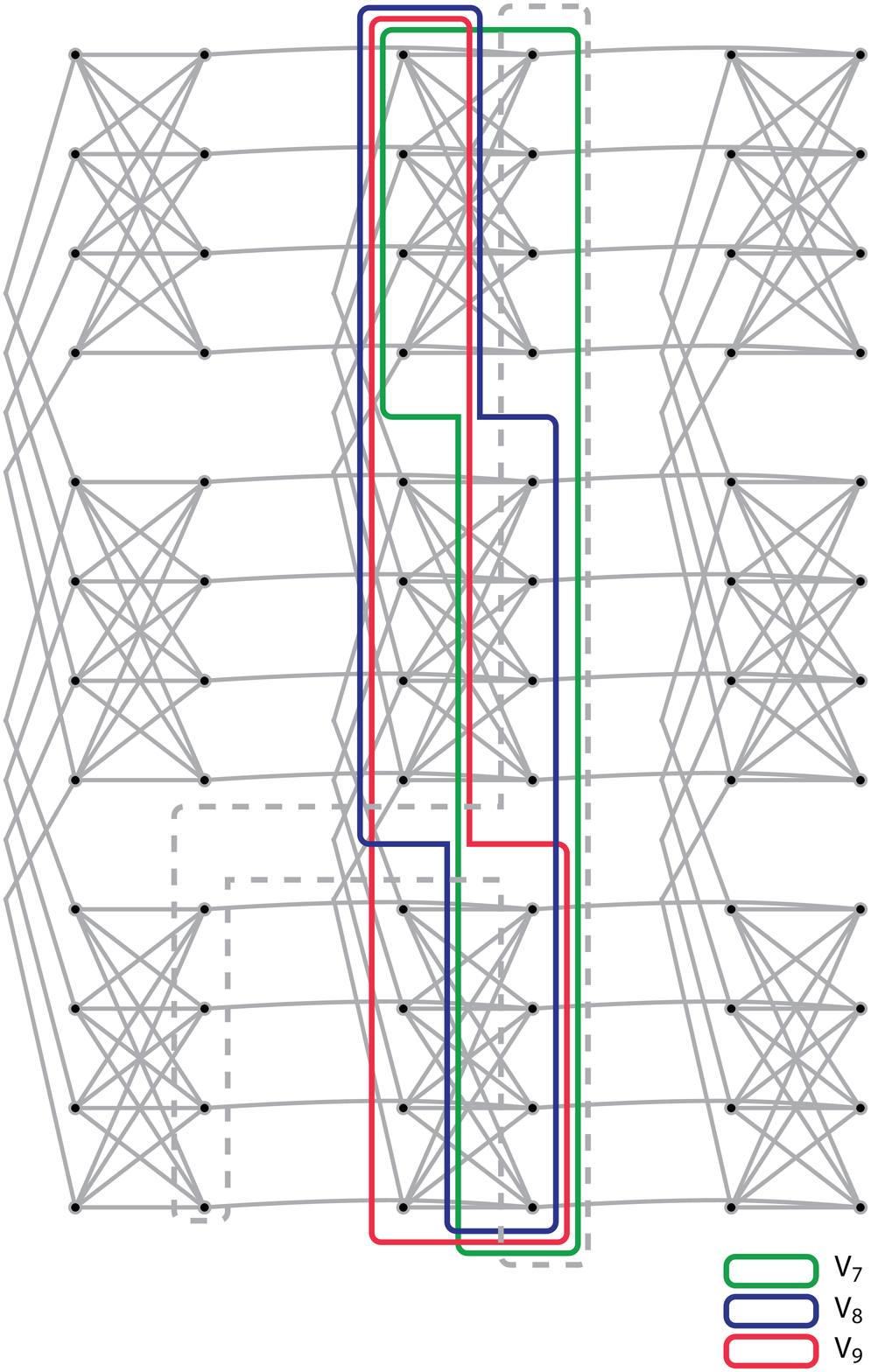}
}
\hspace{0.3in}
\subfigure[]{
	\includegraphics[width=0.302\textwidth]{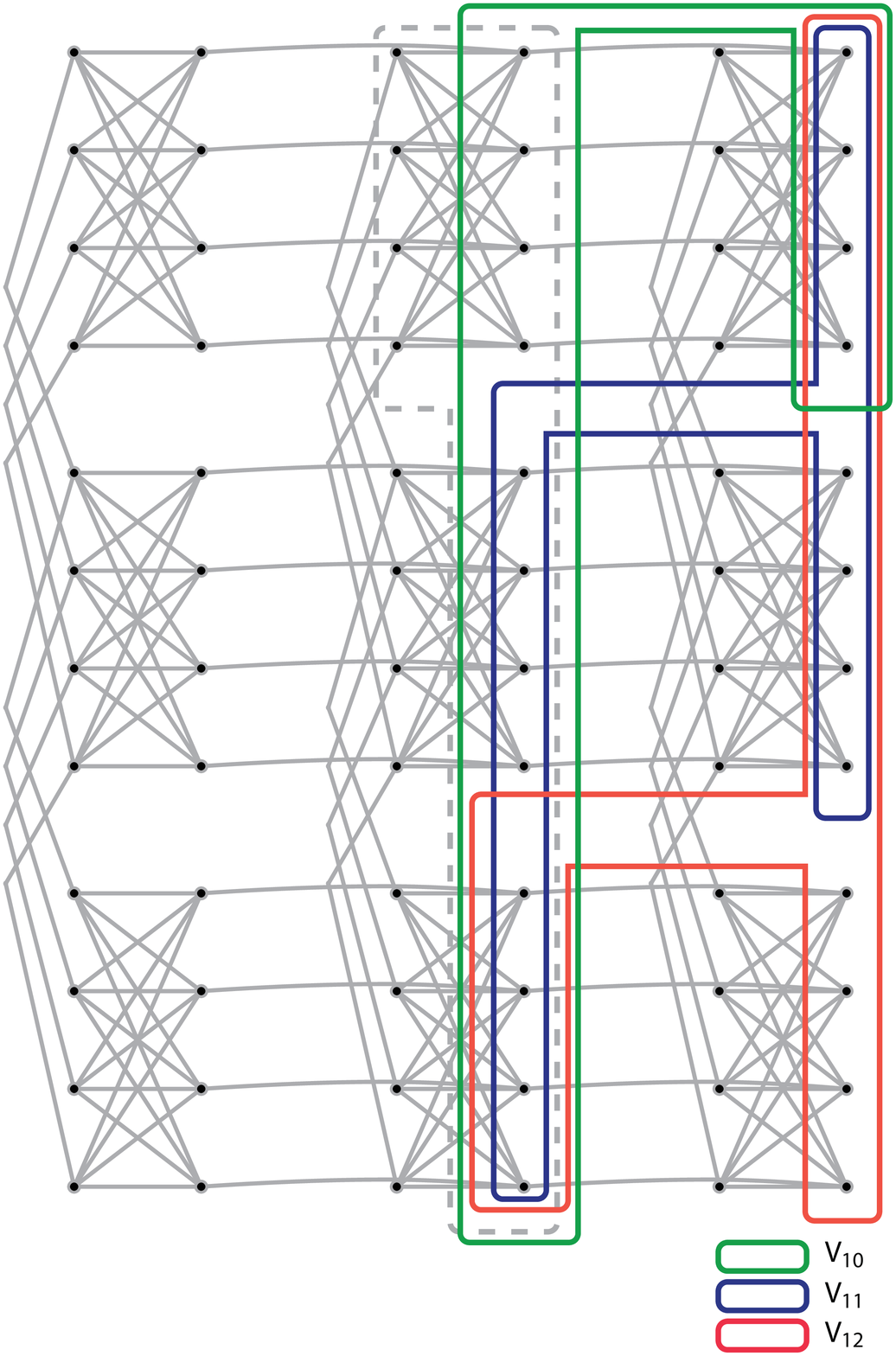}
}
\caption{The first 12 bags of the tree decomposition of a $3\times3$ cell described in the proof of Thm. \ref{thm:hardwaretreewidth}.  The last 3 bags (not shown) have the same layout as the bags in (c), moved to the third column of cells.}
\label{fig:treedecomp}
\end{figure*}

However, in general, determining the treewidth of an arbitrary graph $G$ is {NP}-complete \cite{Bodlaender96,Bodlaender10}.  In \cite{Bodlaender96}, Bodlaender describes a linear time algorithm to determine whether a graph has treewidth at most $k$, for a given fixed $k$.  However, the constants for the algorithm are extremely large (and grow exponentially with $k$), making it impractical for most graphs, including the hardware graphs of interest here.  Amir describes a polynomial-time algorithm which finds a factor-$\mathcal{O}(\log(\tau(G)))$ approximation of the treewidth of a graph $G$ \cite{Amir}, however we have tighter bounds for the treewidth of the hardware graph $F(m,c)$, as presented below. \\

\begin{thm}
\label{thm:hardwaretreewidth}
	Let $F(m,c)$ be a hardware graph made up of an $m\times m$ array of cells, attached as described in Sec. \ref{sec:hardwaregraph}, where each cell contains $2c$ qubits connected to form a $K_{c,c}$.  Then, 
\begin{enumerate}
	\item the treewidth of a single cell ($m=1$) is $c$.
	\item $cm \leq \tau(F(m,c)) \leq cm+c-1$ for $m \geq 2$. \\
\end{enumerate}
\end{thm}

\begin{cor}
\label{cor:Pnotembeddable}
	Any QUBO problem $P$ of treewidth $\tau(P) \geq cm+c$ is not minor embeddable in the hardware graph $F(m,c)$. \\
\end{cor}

\begin{cor}
\label{cor:nobigKn}
	Any QUBO problem which contains a $K_{cm+c+1}$ (either as a subgraph or as a minor) cannot be embedded into  the hardware graph $F(m,c)$. \\
\end{cor}

\begin{cor}
\label{cor:nobiggrid}
	Any QUBO problem which contains a $c(m+1)\times c(m+1)$ grid (either as a subgraph or as a minor) cannot be embedded into the hardware graph $F(m,c)$. \\
\end{cor}

Thus, even though the hardware graph described in Corollary \ref{cor:nobigKn} contains $2cm^2$ qubits, a $K_{cm+c+1}$, which would need only $c(m+1)+1$ logical qubits (if they were all coupled in the fabric), is shown to not be embeddable, due to its treewidth.\\

\begin{IEEEproof}[Proof of Thm. \ref{thm:hardwaretreewidth}]
The proof of (1) follows directly from Lemma \ref{lem:twKn,n}. Furthermore, the lower bound of (2) follows from using the algorithm in Sec. \ref{sec:algorithm} to embed a $K_{cm+1}$ into $G$, since by Lemma \ref{lem:twKn}, $\tau(K_{cm+1})=cm$,  and Lemma \ref{lem:twlowerbound} implies $$cm=\tau(K_{cm+1}) \leq \tau(F(m,c)).$$

	The upper bound is slightly harder to compute.  The proof consists of constructing a tree decomposition of $F(m,c)$ with width $cm+c-1$.  Then, since the treewidth of $F(m,c)$ is the minimum width over all tree decompositions, $cm+c-1$ is an upper bound.

	To form a tree decomposition $(T,\mathcal{V})$ of width $cm+c-1$, we start with $V_1 = \{v_{1,1}^{c+1}, v_{1,1}^{c+2}, \ldots, v_{1,1}^{2c}, v_{1,2}^{c+1},$ $v_{1,2}^{c+2}, \ldots, v_{1,2}^{2c}, v_{1,m}^{c+1}, v_{1,m}^{c+2}, \ldots,v_{1,m}^{2c}, v_{1,1}^{1}, v_{1,1}^{2}, \ldots, v_{1,1}^{c} \}$.  That  is, $V_1$ contains the right half of every cell in the first column of the grid plus the left half of the $(1,1)$ cell.
	
	The idea is to create all other bags of the decomposition by sequentially dropping/adding the left/right halves of individual cells. Each new bag will be formed by removing one of these sets of four vertices from an existing bag, and adding a (different) set of four - specifically one that is not yet contained in any existing bag.  The large amount of overlap between the bags is to ensure that the third requirement of Def. \ref{def:treedecomp} is satisfied.
	
	The bags $V_2, \ldots, V_m$ of the decomposition are formed by dropping the right sides of cells in the first column and picking up the left sides, one-by-one.  That is, $V_i$ contains the right half of cells $i+1$ through $m$  in the first column, the left half of cells 1 through $i-1$, and all of cell $i$.  More formally, for $2 \leq i \leq m$, $V_i = \{v_{i,1}^{c+1}, \ldots, v_{i,1}^{2c},$ $ \ldots, v_{m,1}^{c+1}, \ldots, v_{m,1}^{2c},$ $ \ldots, v_{1,1}^{1}, \ldots , v_{1,1}^{c}, \ldots,$ $ v_{i,1}^{1}, \ldots, v_{i,1}^{c} \}$.  In the tree being formed, $T$, the first $m$ nodes form a path.
	
	The next $m$ bags are formed by (again) starting with $V_1$ but adding the right hand sides of the cells in the second column: for $V_{m+1}$ we drop the remaining four vertices in the left half of the first column and add the top four in the right half of the second; for $V_{m+i}$ with $2 \leq i \leq m$, we add $v_{2,i}^{c+1},\ldots, v_{2,i}^{2c}$, and remove $v_{1,i-1}^{c+1},\ldots,v_{1,i-1}^{2c}$. Bag $V_{2m+1}$ is then formed by dropping the last four vertices from the first column and adding the four left vertices of the top cell in the second column.  Note that $V_{2m+1}$ is the exact same ``shape'' as $V_1$, only one column over.  There is an edge between node 1 and node $m+1$ in $T$, then nodes $m+2$ through $2m$ continue the path.
\par	
At this point, the tree decomposition branches, with two new bags attached to $V_{2m+1}$ (analogous to $V_1$). 
The first is $V_{2m+2}$, which starts the branch consisting of $V_{2m+2}, \ldots, V_{3m}$, with $V_{2m+i}$ dropping $v_{2,i-1}^{c+1}, \ldots v_{2,i-1}^{2c}$ and 
adding $v_{2,i}^{1}, \ldots v_{2,i}^{c}$. Note this is equivalent to how $V_1, \ldots, V_m$ were created. Also attached to $V_{2m+1}$ is $V_{3m+1}$, formed by removing the four righthand vertices from the top cell and adding the top four vertices from the right half of the third column. This branch continues to form 
$V_{3m+2}, \ldots, V_{4m}$ analogously to $V_{m+2}, \ldots, V_{2m}$, so that $V_{4m}$ has the same shape as $V_{2m}$, only one column over.   
\par
The remainder of the tree decomposition is created starting from $V_{4m+1}$ (formed analogously to $V_{2m+1}$), until each column has been covered with a set of bags 
which are formed like $V_1, \ldots, V_m$. This generates a total of $2m^2 - m$ bags, each containing exactly $cm+c$ vertices of $F(m,c)$.  
A small example of the beginning of this process on a $3\times3$ grid of $K_{4,4}$ cells can be seen in Fig. \ref{fig:treedecomp}. The tree associated with this 
tree decomposition can be found in Fig. \ref{subfig:3x3tree}, along with the trees associated with the tree decompositions of the $2\times2$ and the $4\times4$ grids in Fig. \ref{subfig:2x2tree} and \ref{subfig:4x4tree} respectively. Note that these three trees have the same general shape, with only the length of their branches changing, dependent on $m$.
\par	
We now show why $(T,\mathcal{V})$ satisfies the three properties of a tree decomposition from Def. \ref{def:treedecomp}:
\begin{enumerate}
\item every vertex of $F(m,c)$ is in at least one bag.
\item every edge is contained in at least one bag.  This can be verified by noticing that every cell is fully contained in exactly one bag, covering all edges within $K_{c,c}$.  Additionally, for each column, there is a bag containing all of the left side vertices of the cells in the column, and thus all the vertical intercell edges in the column.  Finally, as the bags move from one column to the next, the right halves of each pair of horizontally adjacent cells are contained in a unique bag, thus covering all horizontal intercell edges.
\item Let $v$ be an arbitrary vertex in $F(m,c)$ and let $V_k$ be the lowest index bag in which $v$ appears.  Then, as we walk along $T$ starting at node $k$ and traveling in the direction of increasing node labels, once $v$ is dropped from the bag (on any branch) it is never picked up again.  Thus, the nodes of $T$ which correspond to bags that contain $v$ form a connected subtree of $T$.
\end{enumerate}

\begin{figure}
\centering
\subfigure[]{
	\includegraphics[width=.2\textwidth]{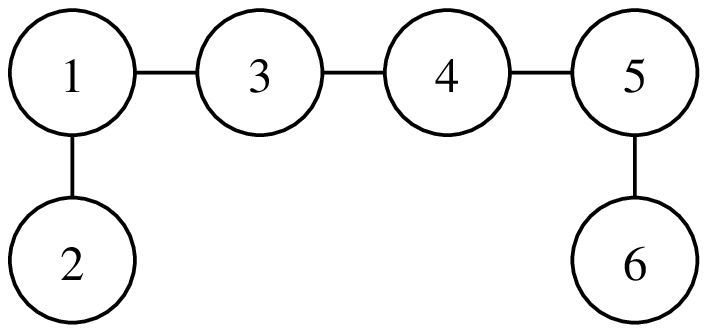}
	\label{subfig:2x2tree}
}
\hspace{0.3in}
\subfigure[]{
	\includegraphics[width=0.23\textwidth]{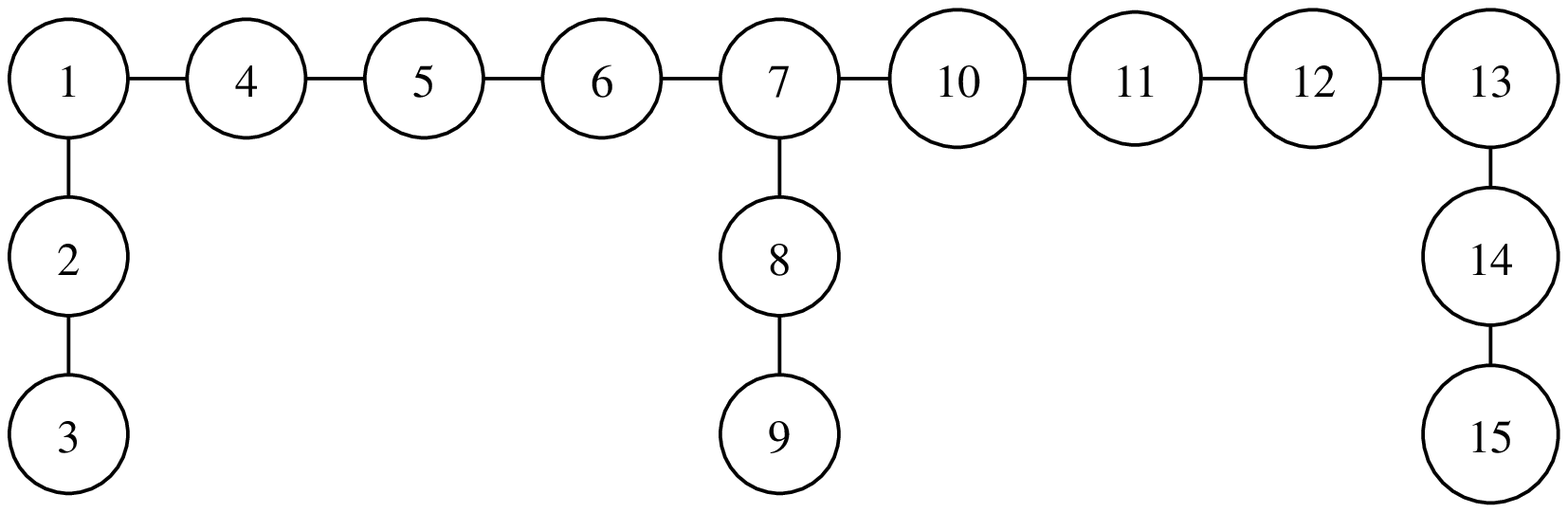}
	\label{subfig:3x3tree}
}

\subfigure[]{
	\includegraphics[width=0.3\textwidth]{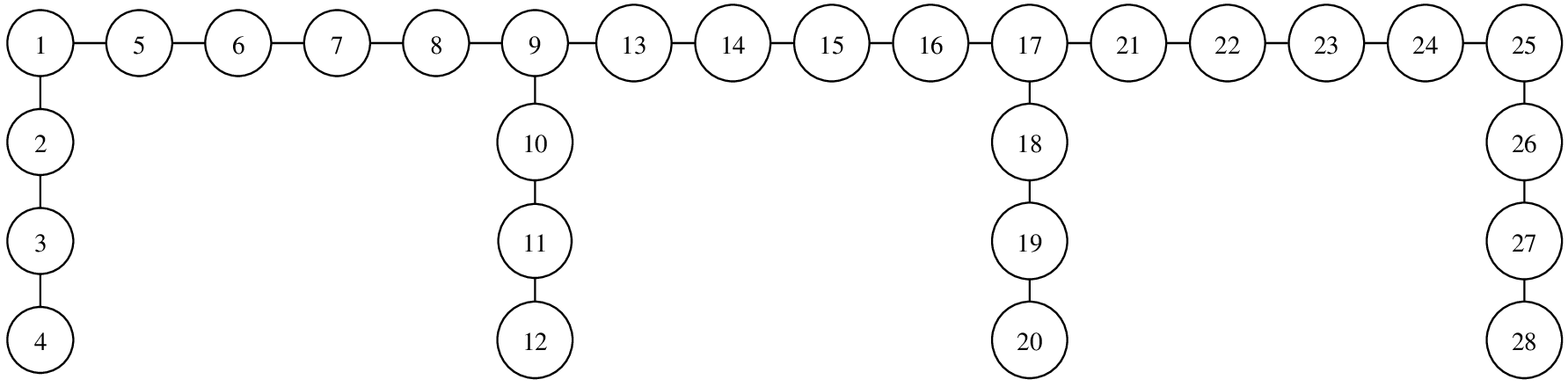}
	\label{subfig:4x4tree}
}
\caption{Trees of the tree decompositions of the hardware graph with a grid of size (a) $2\times2$, (b) $3\times3$, and (c) $4\times4$ which satisfy the upper bound from Thm. \ref{thm:hardwaretreewidth}.}
\label{fig:trees}
\end{figure}

	Since $(T,\mathcal{V})$ is a tree decomposition of $F(m,c)$ where every bag contains $cm+c$ vertices, it has width $cm+c-1$, so $$\tau(F(m,c)) \leq cm+c-1.$$ 
\end{IEEEproof}

While these bounds are not tight for all choices of $c$, they are best possible when $c=1$, as $cm = cm+c-1 =m$.

Determining bounds on the treewidth of the hardware graph is useful because it allows us to automatically dismiss the possibility of embedding certain classes of QUBO problems, members of which we might otherwise have spent considerable time attempting to embed.

If lower bounds on the treewidth of the QUBO problems are known, these can be combined with the bounds on the treewidth of the hardware graph to rule out even more QUBO problems.  There are many graph-theoretic methods for finding lower bounds on treewidth, which use various graph properties including smallest degree, second smallest degree, girth, and spectral radius.  Applying lower bounds to classes of QUBO problems is beyond the scope of this paper, but an overview of common lower-bound algorithms can be found in \cite{Bodlaender10}.

\section{Embedding into the Hardware graph}
In general, determining whether an arbitrary graph $H$ can be minor-embedded into an arbitrary fabric $F$ is NP-complete.  The best-known general algorithms assume a fixed input graph $H$ \cite{Adler}, which is the opposite of the situation in the quantum programming problem.  Additionally, although there are polynomial time recognition algorithms for the existence of an embedding, they do not produce the embedding and, in all cases, the hidden constants are prohibitively large  \cite{Bodlaender96,RobSey}.  Algorithms which allow $H$ to vary along with $F$ are no longer polynomial \cite{Adler,Xiong} or are limited to specific classes of graphs which do not include the hardware graphs described in Sec. \ref{sec:hardwaregraph}  \cite{Kleinberg}.

\subsection{Maximal Minor Embedding}

\begin{figure}[b]
\centering
\subfigure[$K_{4,4}$]{
	\includegraphics[width=.13\textwidth]{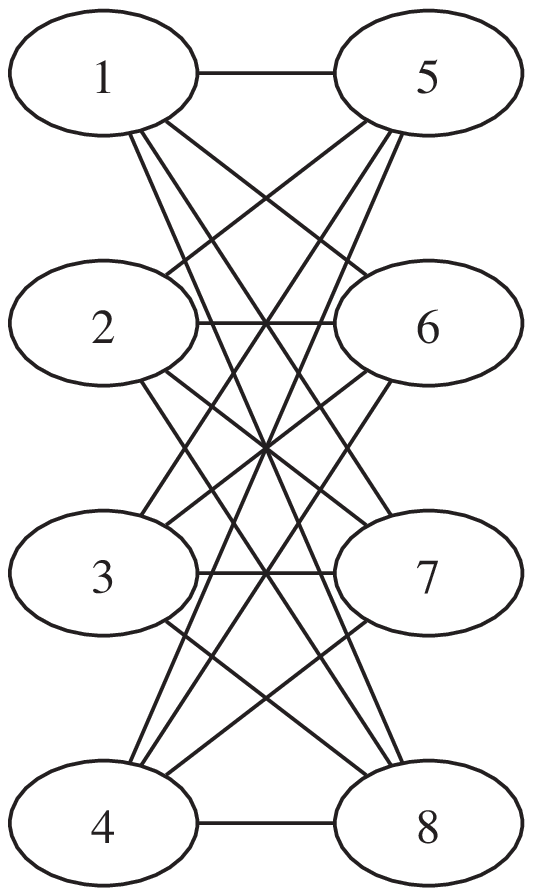}
}
\hspace{0.3in}
\subfigure[$7$-wheel plus $3$ edges]{
	\includegraphics[width=0.23\textwidth]{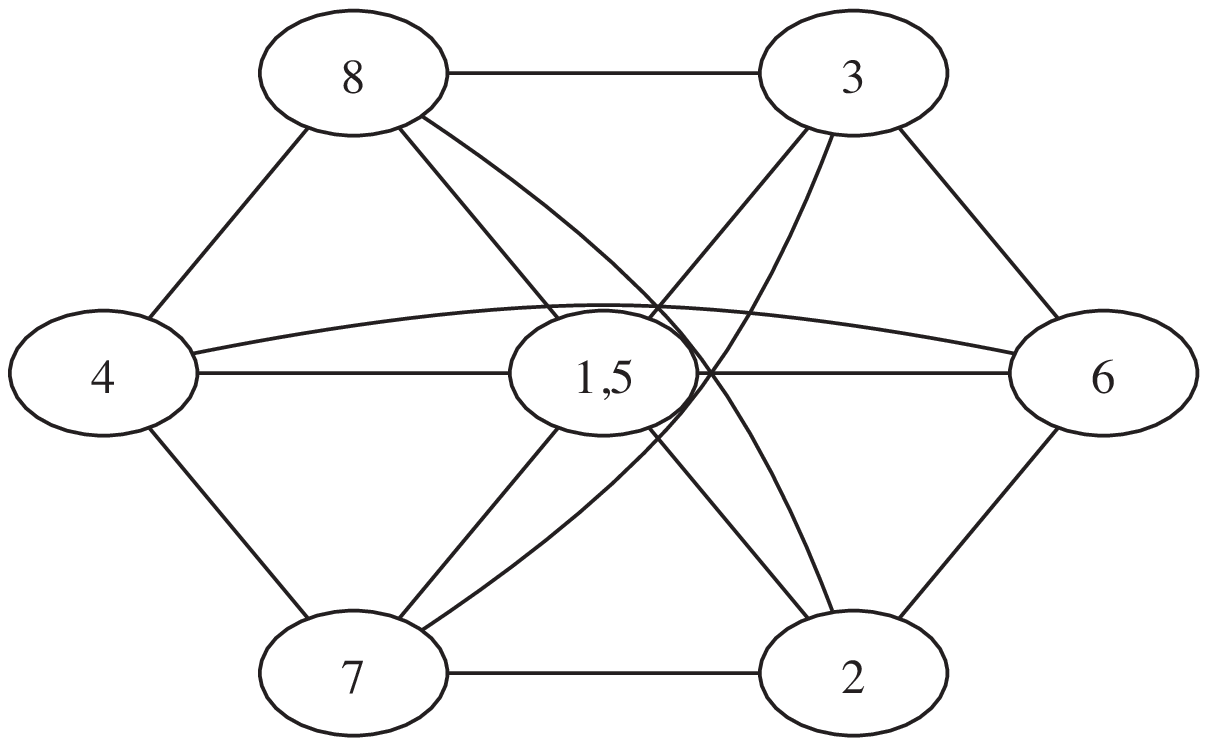}
}

\subfigure[$K_6$ minus $2$ edges]{
	\includegraphics[width=0.18\textwidth]{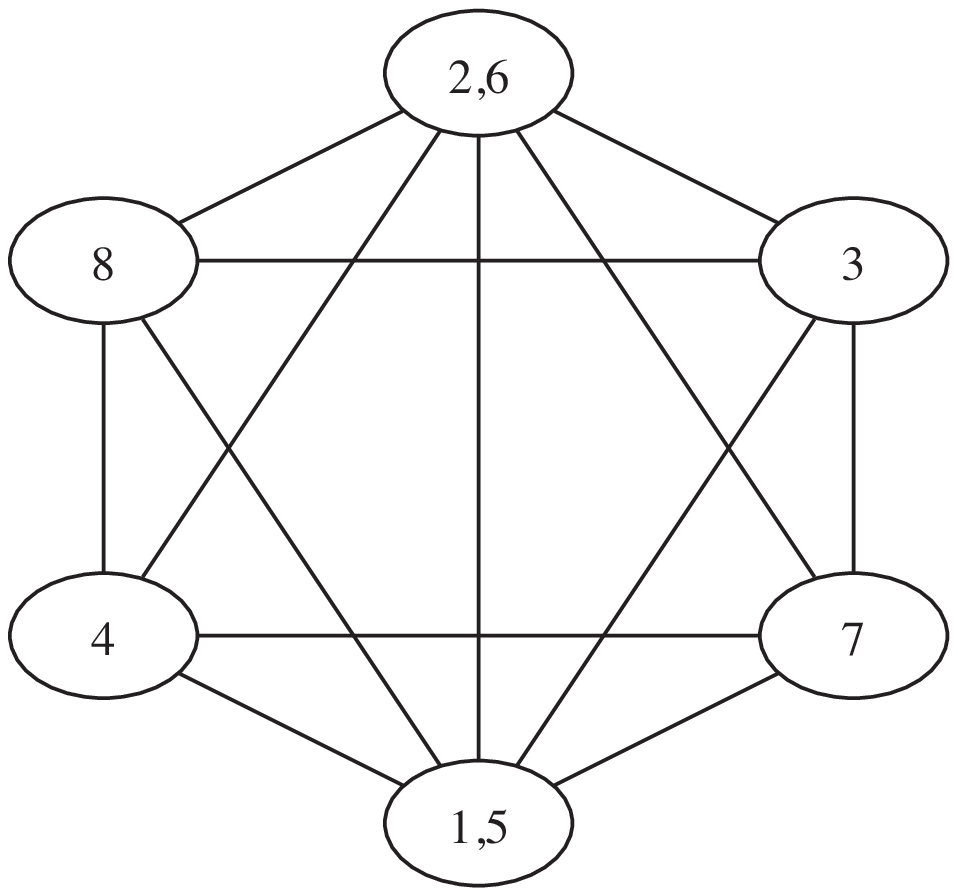}
}
\hspace{0.3in}
\subfigure[$K_5$]{
	\includegraphics[width=0.22\textwidth]{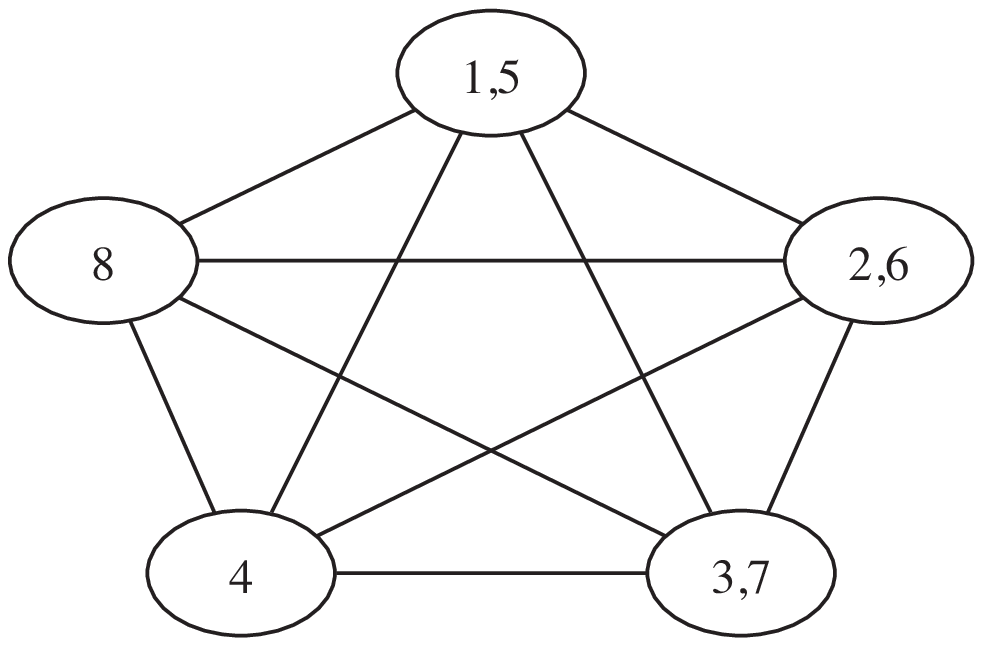}
}
\caption{All minors of a single cell are a subgraph of one of these 4 graphs.}
\label{fig:maxminors}
\end{figure}
\par
Given a fabric $F$ on $n$ vertices, the method for finding and embedding every possible minor-embeddable problem graph $P$ involves solving an NP-complete problem.  First, all the minors of $F$ must be found and, second, we must determine whether $P$ is a subgraph of any of them. The first step can be done when fabric is defined but even once all the minors are known, every new problem graph $P$ must be checked against them for subgraph containment, which is still NP-complete on arbitrary inputs.
\par
The brute force algorithm for finding all possible minors of $F$ involves finding the maximal minors: a set of minors of $F$ such that every other minor is a subgraph of one of the maximal minors.  The first maximal minor is $F$ itself.  Subsequent maximal minors are found by contracting an edge in $F$ to form a minor and checking it for subgraph containment against the list of maximal minors.  If it is not a subgraph of any of these, it is added to the list.  Once every minor of size $n-1$ is found (i.e. every possible edge contraction of $F$ has been tested), the process is repeated by contracting edges in these minors.  The process is completed at step $k$ when no new maximal minors of size $n-k$ are found.  An example of a set of maximal minors can be found in Fig. \ref{fig:maxminors}, which shows the four distinct maximal minors of $F(4,4)$.
\par
Conceptually, maximal minor embedding is very straightforward. The input graph $P$ is compared to the known list of maximal minors for $F$. However, the comparison requires testing for subgraph containment, which is a combinatorial in the number of checks that must be performed. Consequently, maximal minor embedding suffers from two distinct bottlenecks, i.e., finding the maximal minor and finding the embedding. Nonetheless, this method has the benefit of finding the optimal embedding with respect to the size of the embedded problem. Because smaller embedding sizes may be expected to contribute favorably to the scaling of the energy gap, the effort required must be weighed against its advantages.

\begin{figure}
\centering
\subfigure[$K_5$ embedded in a $K_{4,4}$]{
	\includegraphics[width=.16\textwidth]{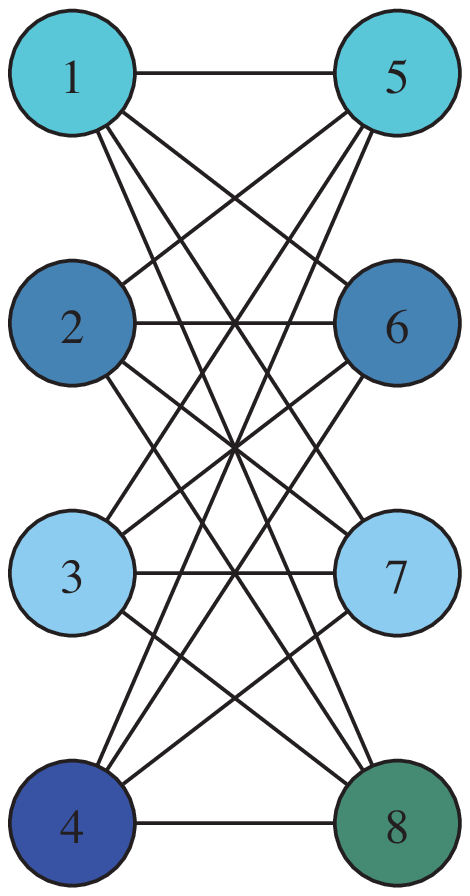}
	\label{subfig:qubitnumbering}
}
\hspace{0.3in}
\subfigure[$K_5$ labelled by hardware vertices merged for each logical qubit]{
	\includegraphics[width=0.23\textwidth]{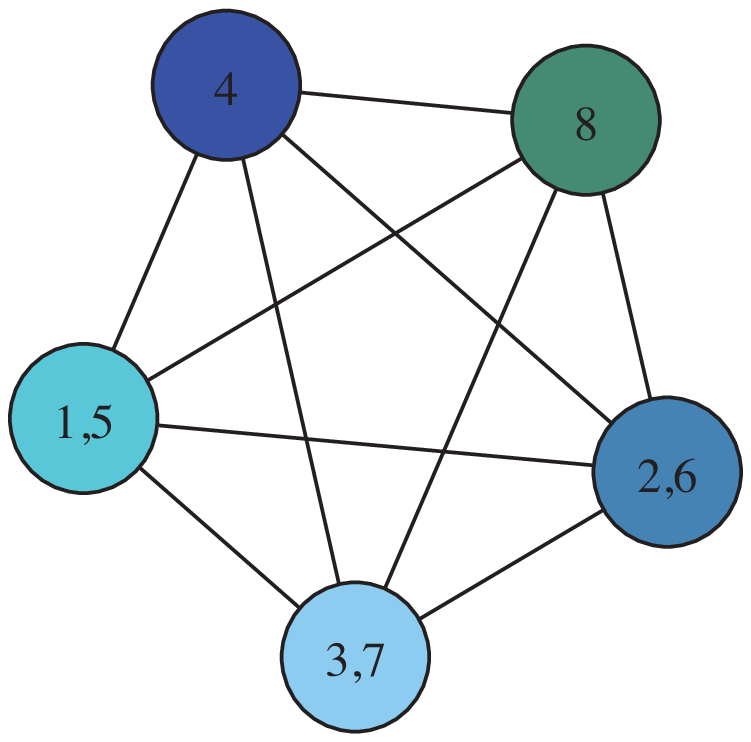}
}
\caption{A $K_5$ embedding into a single $K_{4,4}$ cell of qubits.}
\label{fig:singlecellembedding}
\end{figure}

\subsection{Algorithm to Embed $K_n$}
\label{sec:algorithm}
Instead of trying to find every possible minor of the hardware graph, we can find an embedding of $K_{cm+1}$.  Then, for any QUBO problem of size $cm+1$ or smaller, the embedding problem is solved.  The downside of this approach is that it will fail to embed many problems that are indeed embeddable in the hardware.  For example, although the graphs in Fig. \ref{fig:maxminors}(b) and (c) are embeddable in a $K_{4,4}$ cell, they are not embeddable in $K_5$, which is the largest $K_n$ minor in the cell.  Because of this, the complete-graph embedding algorithm (as described in Sec. \ref{sec:algorithm}) requires a $2\times2$ array of four cells in order to find an embedding for QUBO problems corresponding to either of those graphs.
\par
Unlike maximal minor embedding, the complete-graph embedding algorithm is computationally simple albeit at the cost of increasesd usage of the logical fabric. This illustrates that the two methods described here represent a tradeoff between the computational complexity of the embedding algorithm and the potential computational complexity of the quantum program as measured by the area of the computational fabric.

Given a hardware graph as described in Sec. \ref{sec:hardwaregraph}, our algorithm to embed $K_{cm+1}$ as a minor in the $m \times m$  grid of $K_{c,c}$ cells is recursive in nature, and constructs the mapping $\phi$ described in Def.~\ref{def:minor}.   For the sake of clarity, in the description of the algorithm, the elements of the $K_{cm+1}$ will be referred to as nodes and the elements of the hardware graph will be referred to as vertices.  Let $u_1, u_2, \ldots, u_{cm+1}$ be the nodes of the $K_{cm+1}$ that we are trying to embed.  

The algorithm begins by embedding the first $c+1$ nodes (forming a $K_{c+1}$) into the cell in the upper left corner of the hardware.  This is done by pairing left and right vertices $c-1$ times. 
\newpage
\begin{lstlisting}[language=Matlab]
function V = no_failure_embedding(c,m)
% This function takes an mxm hardware graph 
% of K_{c,c} cells and outputs a (2m)x(cm+1) 
% matrix V where the non-zero entries of V(:,i) 
% are phi(u_i) for u_i in the embedded K_{cm+1}

V = zeros(2*m,c*m+1);
%Almost all the c*m-2 sets are formed similarly
for i = 1:c*m+1
    if i < c
        %First grid row/column, position i in cell
        r = 1; s = i;
    else if i > c+1
        %Calculate row/column of the grid 
        r = ceil((i-1)/c);
        %Calculate level within cell   
        s = mod((i-1),c);
        if s==0
            s=c;
    else
        continue; %these are handled below

    %fill in the horizontal members of phi(u_i) 
    for j=1:m
        V(j,i)=2*c*m*(r-1)+2*c*(j-1)+c+s;
    %fill in the vertical members of phi(u_i)
    for j=1:m
        V(j+m,i)=2*c*(r-1)+2*c*m*(j-1)+s;
end
%At i=c and i=c+1, the sets differ, and have size m 
for j=1:m
    V(j,c)= c + (j-1)*2*c*m;
    V(j,c+1)= j*2*c;

\end{lstlisting}

That is, for $1 \leq j \leq c-1$, $\phi(u_j) = \{v_{1,1}^j, v_{1,1}^{c+j} \}$.  The next two nodes are each initially mapped to a set containing a single vertex: $\phi(u_{c}) = \{v_{1,1}^c\}$, and $\phi(u_{c+1}) = \{v_{1,1}^{2c}\}$.  See Fig. \ref{fig:singlecellembedding} for an example of embedding $K_5$ in a $K_{4,4}$ cell of 8 qubits. We provide an instance of this algorithm in a Matlab-style pseudocode for a function which produces an embedding into non-faulty $F(m,c)$ hardware.

\begin{figure*}
\centering
\subfigure[$K_{13}$ embedded in upper $3\times3$ sub-grid]{
	\includegraphics[width=.443\textwidth]{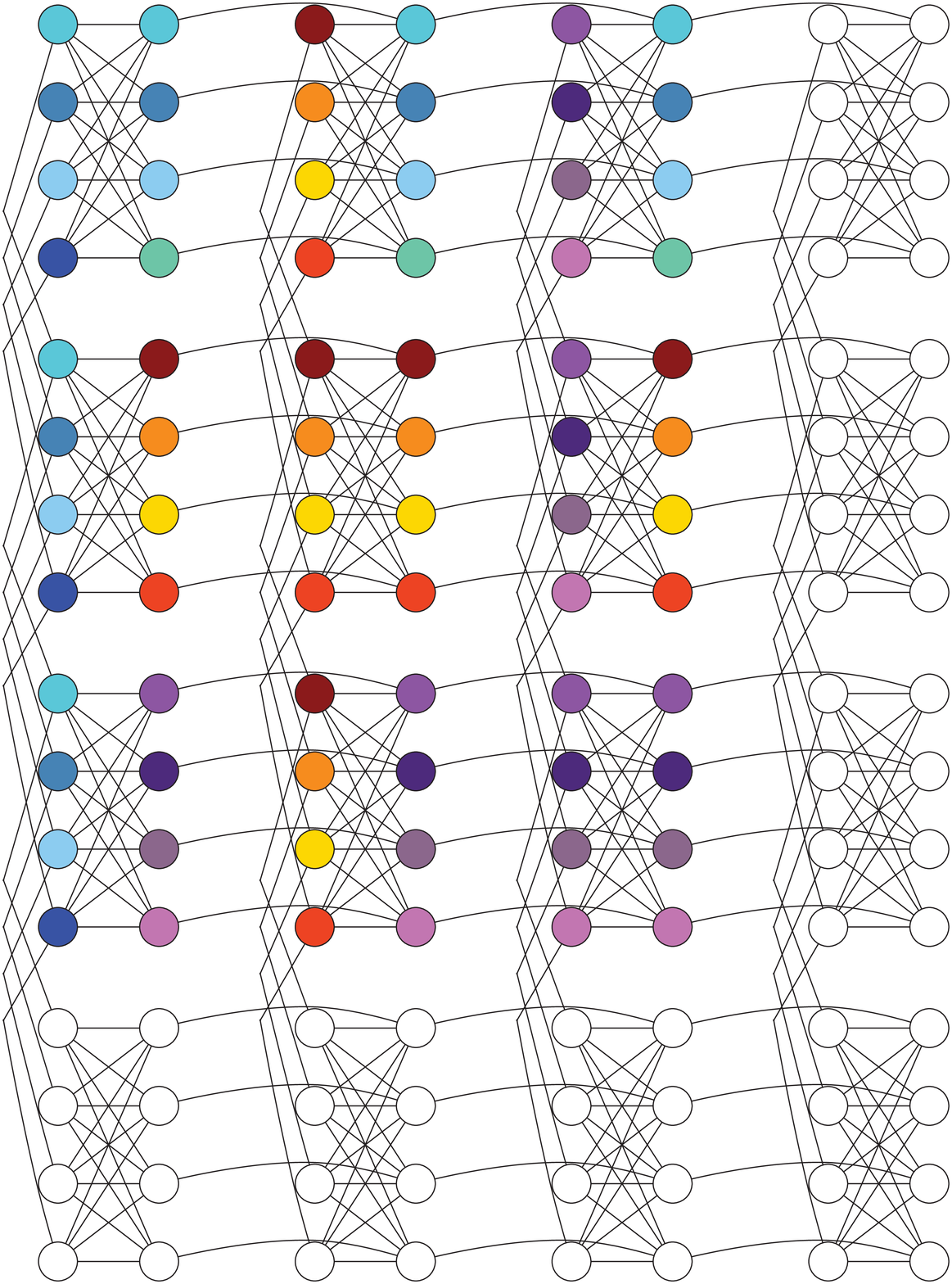}
}
\hspace{0.3in}
\subfigure[$K_{17}$ embedded in the $4\times4$ grid]{
	\includegraphics[width=0.443\textwidth]{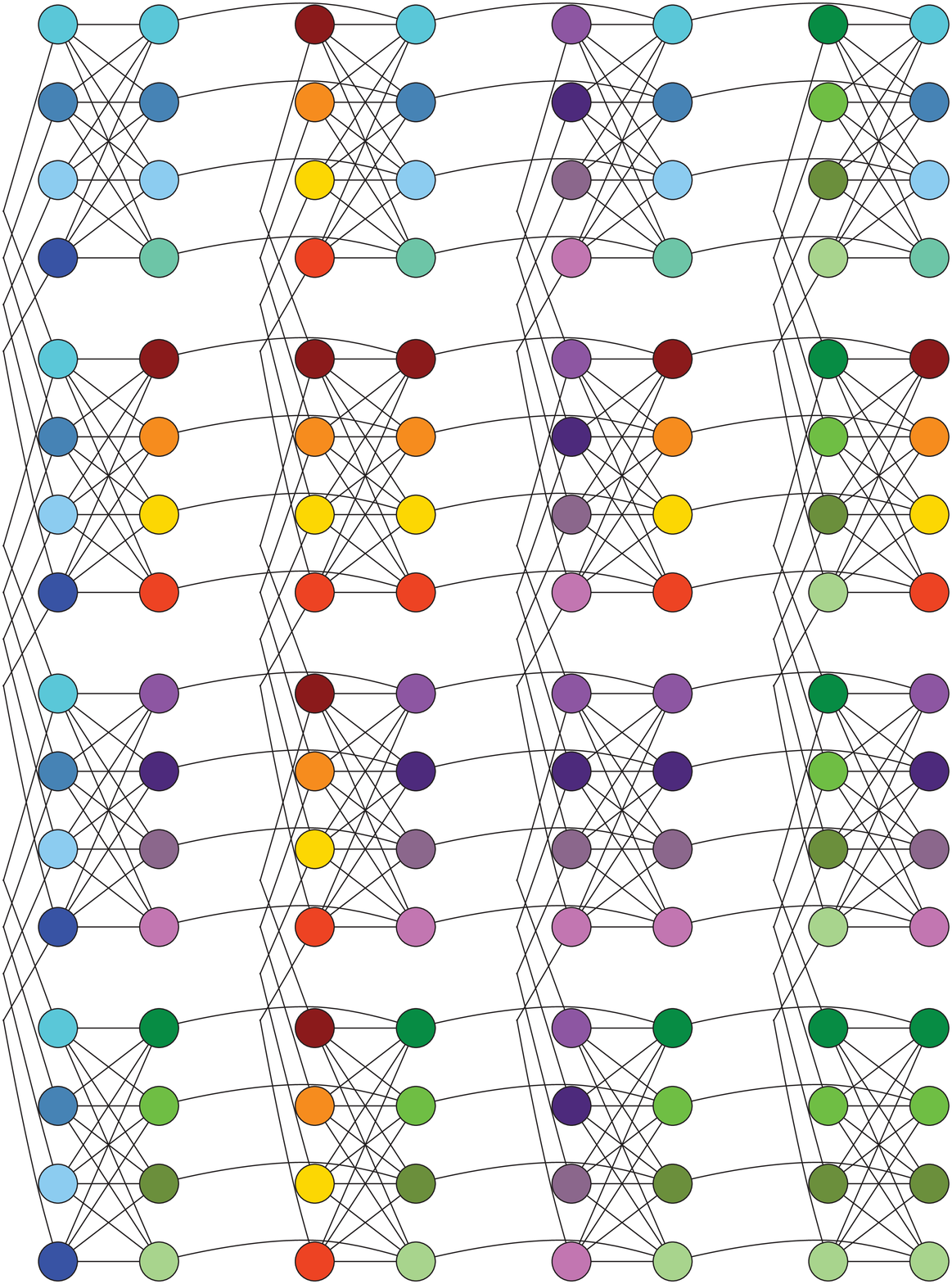}
}
\caption{A $K_{13}$ embedding in a $3\times3$ grid of $K_{4,4}$ cells extended to a $K_{17}$ embedding in a $4\times4$ grid.  Each color represents a single logical qubit.}
\label{fig:Knembedding}
\end{figure*}
\vskip1em
After embedding a $K_{c+1}$ into the first cell of the hardware graph, the $m-1$ remaining steps of the algorithm extend the embedding into the subsequent row and column of the $m\times m$ grid.  For each step $2 \leq i \leq m$, the embedding forms an {\em extendable} clique minor in the $i \times i $ grid.  We say a minor is extendable if it satisfies two conditions: first for $u_j$, $1 \leq j \leq c(i-1)+1$ the set $\phi(u_j)$ is non-empty. Second, each set has at least one vertex with an edge into the next row or column.

For all nodes $u_j$, at least one vertex of $\phi(u_j)$ is connected to a cell in the next row and/or column of the grid.  These vertices are added to the set $\phi(u_j)$.  For nodes $u_c$ and $u_{c+1}$, one vertex is added to $\phi(u_c)$ and $\phi(u_{c+1})$ at each layer $i$.  For all other nodes $u_j$, two new vertices are added to $\phi(u_j)$.

The sets $\phi(u_{c(i-1)+2})$ through $\phi(u_{c(i-1)+c+1})$ are formed by picking one of the unclaimed vertices on the right side of cell $(1, i)$.  This is extended by following the edges from cell to cell along row $i$.  When column $i$ is reached, one edge is taken within the cell, then edges from cell to cell are followed up along column $i$.  At the end of this process, each of these sets will contain $2i$ vertices: for $1 \leq s \leq c$, $\phi(u_{c(i-1)+s+1}) = \{v_{1,i}^{c+s}, \ldots v_{i,i}^{c+s}, v_{i,1}^s, \ldots, v_{i,i}^s\}$.  

This process is continued until $K_{cm+1}$ is fully embedded in the $m\times m$ grid.  See Fig. \ref{fig:Knembedding}  for an extension of a $K_{13}$ embedding in a $3\times3$ grid of $K_{4,4}$ cells to a $K_{17}$ embedding in a $4\times4$ grid of cells.

In the description of the hardware graph in Sec. \ref{sec:hardwaregraph}, the vertices were given labels of the form $v_{a,b}^d$.  In the Matlab-style pseudocode found below, they are numbered from 1 to $cm^2$.  The numbering starts in the cell in the upper left corner as described in Fig \ref{subfig:qubitnumbering} and this numbering is continued across the row, then across subsequent rows.  Given a node position in the form $v_{a,b}^d$, the equivalent number in the code below is $n = 2cm(a-1)+2c(b-1)+d$.  Given a node numbered $n$ in the code below, the equivalent label is given by $v_{a,b}^d$ with $a=\lceil \frac{n}{2cm}\rceil$, $b = \lceil \frac{n-2cm(a-1)}{2c} \rceil$, and $d= n\mod 2c$, with $d=2c$ if $n\mod 2c =0$.

\section{Embedding with failed Qubits}
\label{sec:failures}
The complete-graph embedding algorithm presented in Sec. \ref{sec:algorithm} assumed that there are no failures in the hardware.  However, the hardware may exhibit some percentage of failed vertices which prevent a full $K_{cm+1}$ embedding (e.g. in the case of any single qubit failure, the biggest clique embeddable is $K_{cm}$).  Instead of losing a node from the $K_{cm+1}$ for each failed qubit, techniques can be employed to embed in a way that attempts to minimize the number of sets $\phi(u)$ which contain any failed qubits. 
\par
We present two algorithms below in order to handle the case of fabrics with hard faults. These approaches to embedding test the different starting points available from the four corners of the $m\times m$ grid and then return the best possible embedding that results.  
Additionally, if the largest $K_n$ found is smaller than the largest possible in an $(m-1)\times(m-1)$ grid, from each corner, we drop the first row and column and reattempt the embedding.  This ``dropping down" procedure continues until a large enough clique is found or $(m-1)$ rows and columns have been dropped.  
\par
At the same time, the grid is scanned and the largest $K_n$ embeddable in a single cell ($1\leq n \leq c+1$) is found.  If a complete cell is found, this is $K_{c+1}$.  The reported largest embeddable $K_n$ output by the algorithm is the maximum of the largest clique embeddable inside a single cell and the four cliques found from starting at the four corners.  
\par
Combining these two procedures yields a ``flip and drop-down'' method that we compare to the single, nominal attempt at embedding, i.e., starting in the upper left corner.  In all cases, the worst performance possible is to embed a $K_1$, since we assume there is at least one working qubit in the hardware.  Note details of the corner selection and drop-down methods are not shown in the pseudocode.

\subsection{Dropping to a smaller cell-graph}
\label{sec:fallback}
Given an $m\times m$ hardware graph with cells of $K_{c,c}$, one way to deal with failed qubits is to find the largest $c_o$, $c_o \leq c$, such that there is a complete $m\times m$ grid of $K_{c_o,c_o}$'s and use the algorithm described in Section~\ref{sec:algorithm} to embed into this sub-grid.  This will lead to an embedding of size $c_om+1 \leq cm+1$. Once the $c_o$ has been determined, the embedding can be found by renumbering the vertices of the hardware graph to reflect the new cell size and running no\_failure\_embedding($c_o,m$).

\subsection{Greedy failure algorithm}
\label{sec:greedy}
As can be seen in Fig. \ref{fig:Knembedding}, given a perfect $m\times m$ grid of $K_{c,c}$ cells, for each node $u$ of the embedded $K_{cm+1}$ (other than nodes $u_c$ and $u_{c+1}$ started in the first cell), $\phi(u)$ contains $2m$ vertices.  These consist of two sets of $m$ vertices: a connected set consisting of one vertex from the left side of each cell in the a single column in the grid and a connected set consisting of one vertex from the right side of each cell in the row of the same number.  Due to the pattern in which cells are connected, within both of these sets every vertex occupies the same position in the cell it comes from.

The greedy failure algorithm works to maximize the size of the complete $K_n$ which can be embedded in the hardware graph with failed vertices, by attempting to pair up sets containing failed vertices with other sets containing failed vertices to create full nodes.  These ``match-ups'' occur in the diagonal cells of the grid.  In the case of no failures, each horizontal set (of vertices from the right halves of cells)  is matched with a vertical set (of vertices from the left halves of cells) whose vertices occupy the same `height' inside a single cell.  When there are errors, however, horizontal sets containing failed vertices attempt to match with vertical sets that also contain failed vertices, regardless of the 'heights' at which the vertices sit inside a cell.   By matching sets which contain failures, the number of complete nodes (all of which except  $u_c$ and $u_{c+1}$ are made up of  two sets) containing failures is reduced and, consequently, a larger embedded $K_n$ is achieved. 

The Matlab-style pseudocode for a function which produces the nodes of the embedding described above and outputs the number of nodes containing no errors can be found at right.

\subsection{Analysis}

A comprehensive set of experiments were run to see how well the fallback and greedy algorithms from Secs. \ref{sec:fallback} and \ref{sec:greedy}, respectively, performed under various conditions of vertex failure.  These experiments were run using a single attempt at embedding that begins in the upper left corner of the grid of cells as well as a run using the flip and drop-down scheme described at the beginning of Sec. \ref{sec:failures}.  In all cases, the hardware graph was an $m\times m$ grid of $K_{4,4}$ cells.  The grid sizes tested were $m=4, 8, 16,$ and 32.  For each of these grid sizes, the algorithms were run with a percentage of failed vertices of $p=2,4,5,6,8,10,15, 20\text{ and } 25$.  The failed vertices were uniformly distributed across the hardware graph. In each of the 148 cases (defined by algorithm, scheme, grid size, and failure rate), 10,000 randomized instances were run to compute statistical averages.

\begin{figure*}[ht]
\centering
\subfigure[Single attempt at embedding]{
	\includegraphics[width=.45946\textwidth]{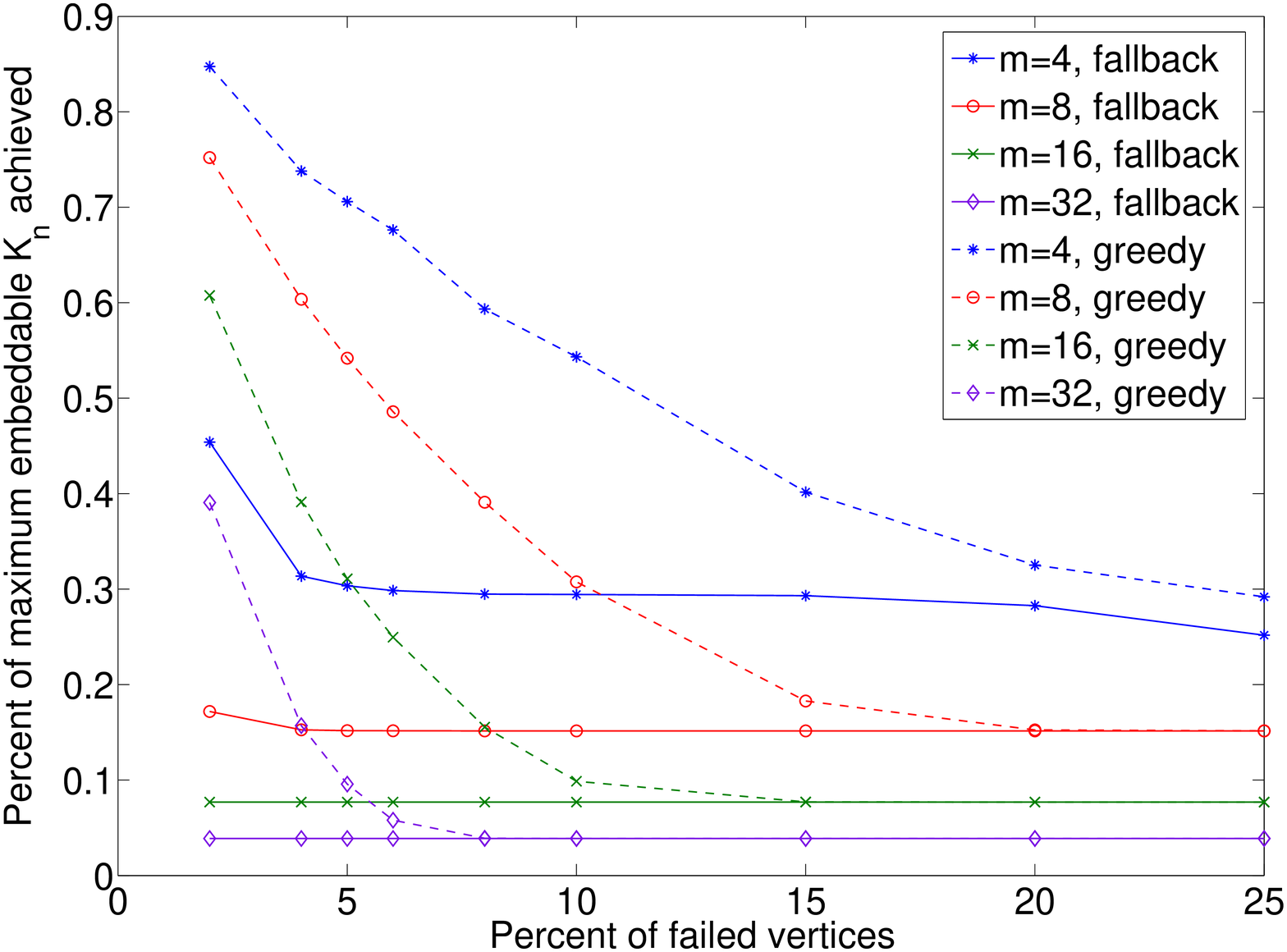}
}
\hspace{0.3in}
\subfigure[Maximum over flip and drop-down embeddings]{
	\includegraphics[width=0.45946\textwidth]{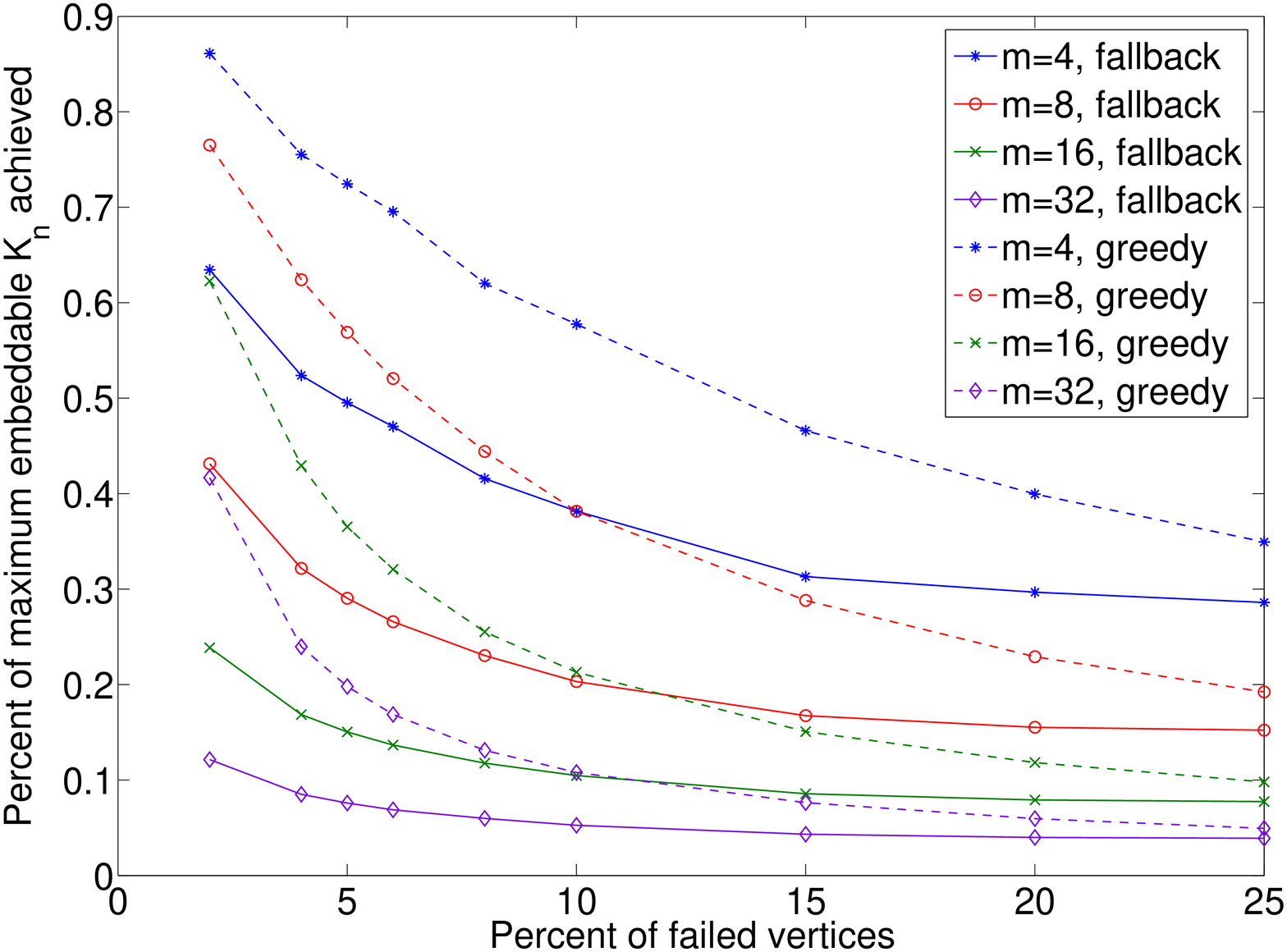}
}
\caption{Percent of the maximum embeddable $K_n$ achieved for both the fallback and greedy embedding schemes for various percentages of failed vertices, averaged over 10,000 trials.  This is calculated for both a single attempt at the embedding (left) and multiple attempts at the embedding, starting in all four corners and, if necessary, dropping to a smaller grid (right).  Both methods also search for a whole cell.}
\label{fig:singlevdropdown}
\end{figure*}

\par
\begin{figure*}[th!]
\centering
\subfigure[Fallback embedding algorithm]{
	\includegraphics[width=.4599\textwidth]{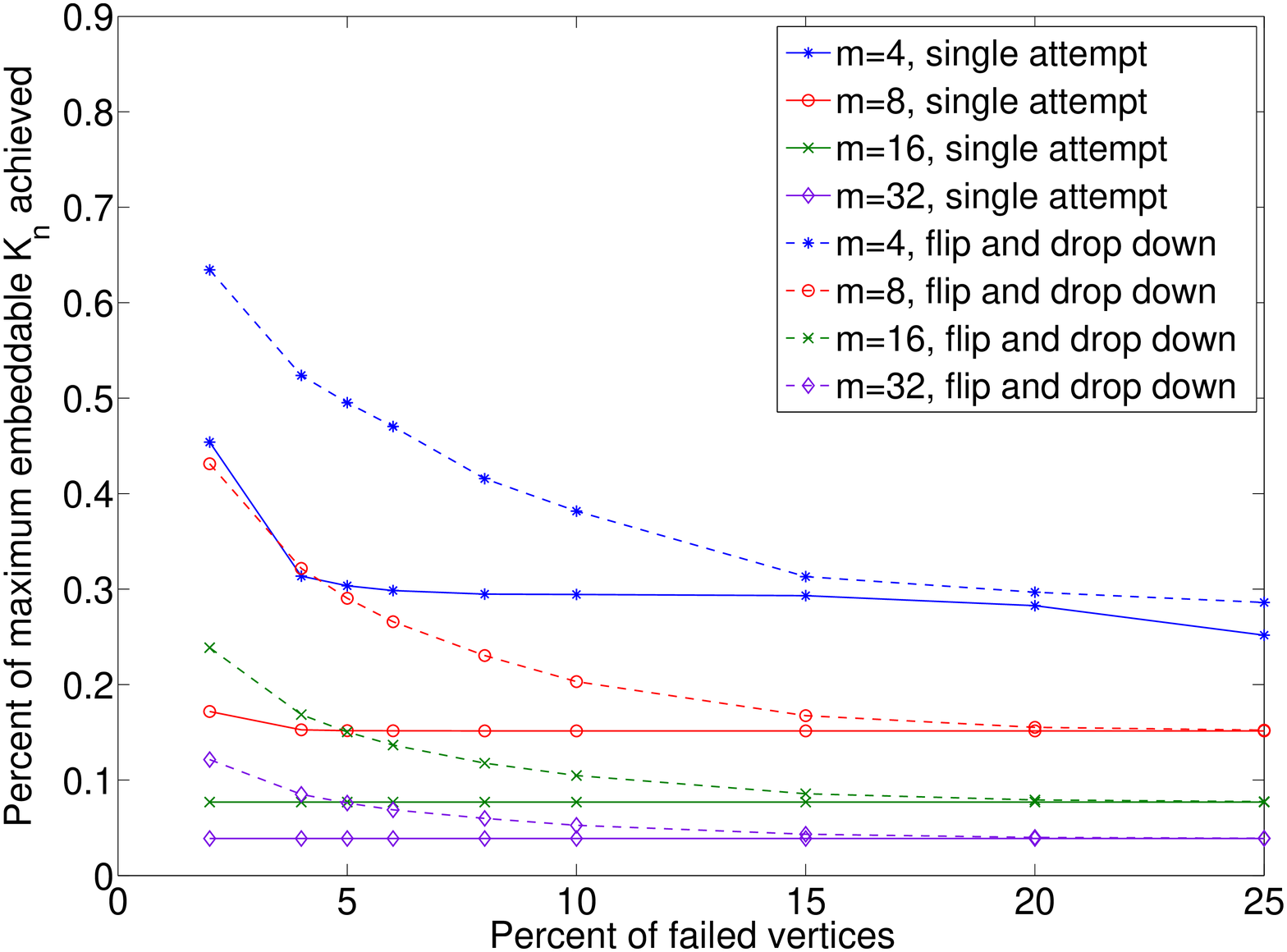}
}
\hspace{0.3in}
\subfigure[Greedy embedding algorithm]{
	\includegraphics[width=0.458\textwidth]{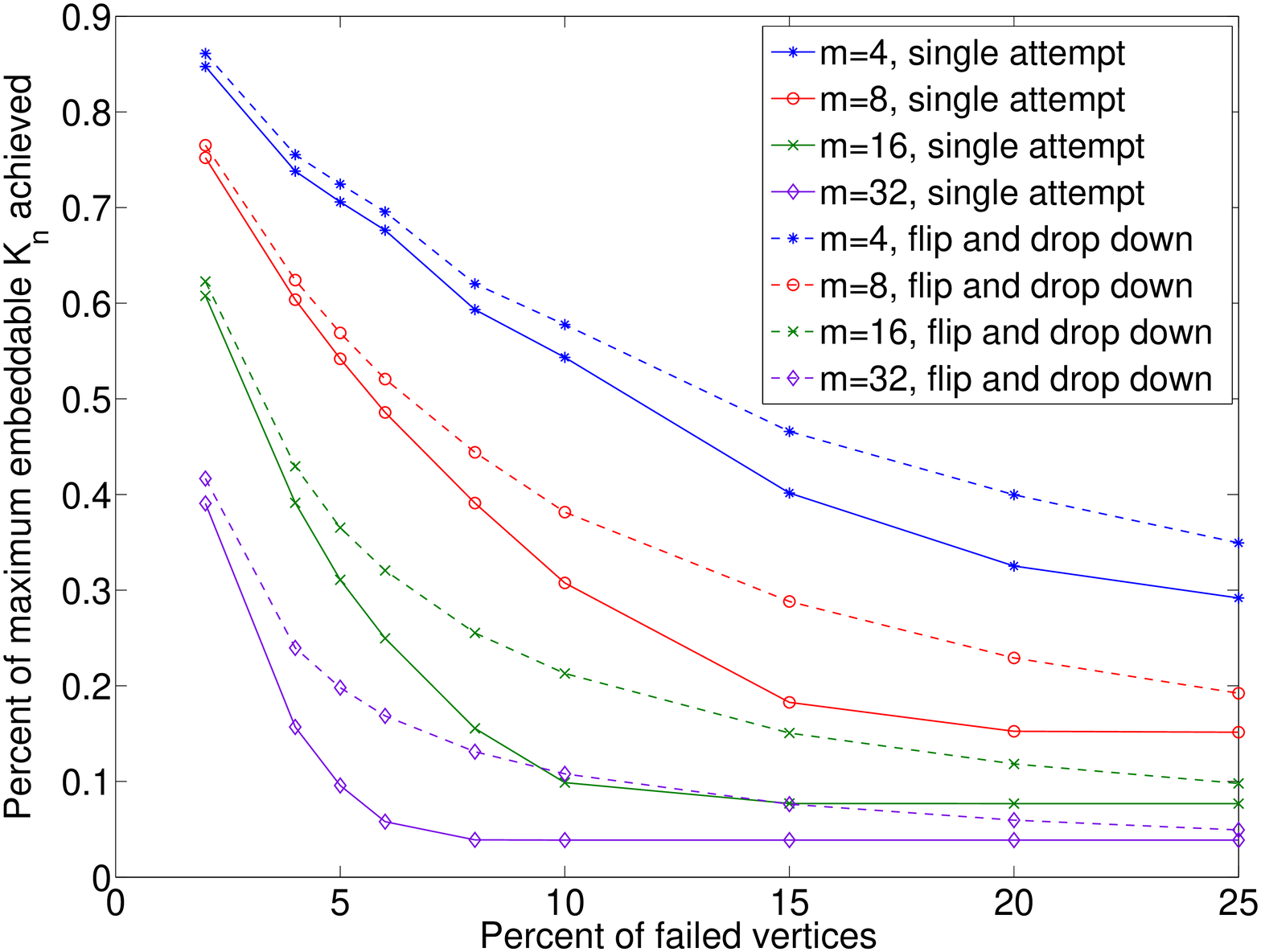}
}
\caption{Percent of the maximum embeddable $K_n$ achieved in both a single attempt at embedding and multiple attempts (starting in each of the four corners and, if necessary, dropping down) for various percentages of failed vertices, averaged over 10,000 trials.  This is calculated both for the fallback method (left) and for the greedy method (right).}
\label{fig:fallbackvgreedy}
\end{figure*}

A comparison of the results shown by Figs. \ref{fig:singlevdropdown} and \ref{fig:fallbackvgreedy} illustrates that the flip and drop-down embedding scheme performs better than a single attempt at embedding from the upper left corner and that the greedy algorithm performs better than the fallback method.  In both schemes, the greedy algorithm embeds a $K_n$ with $n$ approximately 85\% of the optimum value at two percent failure rate.

\newpage
\begin{flushleft}
\begin{lstlisting}[language=Matlab]
function [V,k] = greedy_embedding(c,m,G)
% This function takes an mxm hardware graph 
% of K_{c,c} cells and a list G of failed 
% vertices. Outputs are a (2m)x(cm+1) matrix V, 
% where non-zero entries of V(:,i) are phi(u_i) 
% for u_i in the embedded K_{cm+1}, and k is the 
% number of failure-free sets phi(u_i).

%Helper Function: PAIR(s,t,cv) 
%stores the union of F(:,s) and F(:,t) in V(:,cv)

%First, we form all of the half-sets in a matrix F
F = zeros(m,2*c*m)
for i=1:m
	for pos=1:c
		%determine columns of F to be filled
		Cnum = 2*c*(i-1)+pos
		Rnum = 2*c*(i-1)+pos+c
		for j=1:m
		    %half-sets in col i of hardware graph 
		    F(j,Cnum)=2*c*m*(i-1)+2*c*(j-1)+pos+c
		    %half-sets in row i of hardware graph
		    F(j,Rnum)=2*c*m*(j-1)+2*c*(i-1)+pos
end %of for i=1:m

% Match half-sets for each row/column to minimize 
% number of full sets containing failed vertices.
V=zeros(2*m, c*m+1)
cv = 1; %first open column of V
k=0; %number of failure-free full sets created

for i=1:m
    Fi= 2c*(i-1) %offset for column indices in F
    %Pair up sets containing failures
    for s=1:c
        if i==1 and cv==c
            break; %go create size m sets
	    if F(:,Fi+s) contains a failure in G
	        for t=1:c
	            if F(:,Fi+c+t) contains a failure
	      	        PAIR(s,c+t,cv)
	                cv++
	                break
    end %of for s=1:c

    %Pair remaining half-sets arbitrarily until  
    %c-1 (i=1) or c (i>1) whole sets have been made
    for s=1:c
        if ((i==1 and cv==c) or (cv==c*i+2))  
            break; %create size m sets or next i
        if F(:,Fi+s) unpaired
            for t=1:c
	        if F(:,Fi+c+t) unpaired
	            PAIR(s,c+t,cv)
	            cv++
	            if V(:,cv) failure-free
	                k=k+1
    end %of for s=1:c        
    
    %Create two size m sets in row/column 1:
    for s=1:c
    	if F(:,s) unpaired
            V(:,cv) <- F(:,s) 
            cv++
            if F(:,s) failure-free
                k=k+1
        if(F:,c+s) unpaired
            V(:,cv) <- F(:,c+s) 
            cv++;
            if F(:,c+s) failure-free
                k=k+1
    end %of for s=1:c
end %of for i=1:m
        
        
\end{lstlisting}
\end{flushleft}
\pagebreak 

\par
At fixed failure rate, the percent of the maximum embeddable $K_n$ for both algorithms decreases as the grid size $m$ grows.  This is due to the fact that the number of hardware vertices mapped to a single node of the $K_n$ minor increases linearly with grid size. On the $4\times4$ grid, each set $\phi(u)$ is made up of 8 vertices (except for 2 special cases). Given a 2\% failure rate, this means that any $\phi(u)$ on the $4\times4$ grid (with no attempt at a `smart' embedding scheme) has a 16\% chance that the set contains at least one failed vertex (and thus can not augment the size of the $K_n$ embedded). Similarly, on the $32\times32$ grid, each $\phi(u)$ contains 64 vertices, and for 2\% failure having at least one failed vertex per cell is highly likely.  
\par
At 2\% failure rate, the greedy embedding scheme with flips and drop-downs achieves embedding of a complete graph of over 40\% the size of the maximum $K_n$ embeddable.  For the worst case scenario, and with no attempt at a `smart' embedding, it would only take one failed vertex to destroy each logical qubit.  Even at only a 2\% failure rate, the $32\times32$ grid has on average 163 failed vertices.  If the algorithm did not adapt, this high failure density would completely destroy the maximum embeddable clique, which is a $K_{129}$.  In the case of a 25\% failure rate, the number of failed vertices jumps to 2048, yet the greedy failure algorithm is still able to embed a $K_{6}$ on average.

\begin{figure}[b]
\centering
\hspace{-0.2in}
\subfigure[Fallback algorithm]{
	\includegraphics[width=0.24\textwidth,height=.15\textwidth]{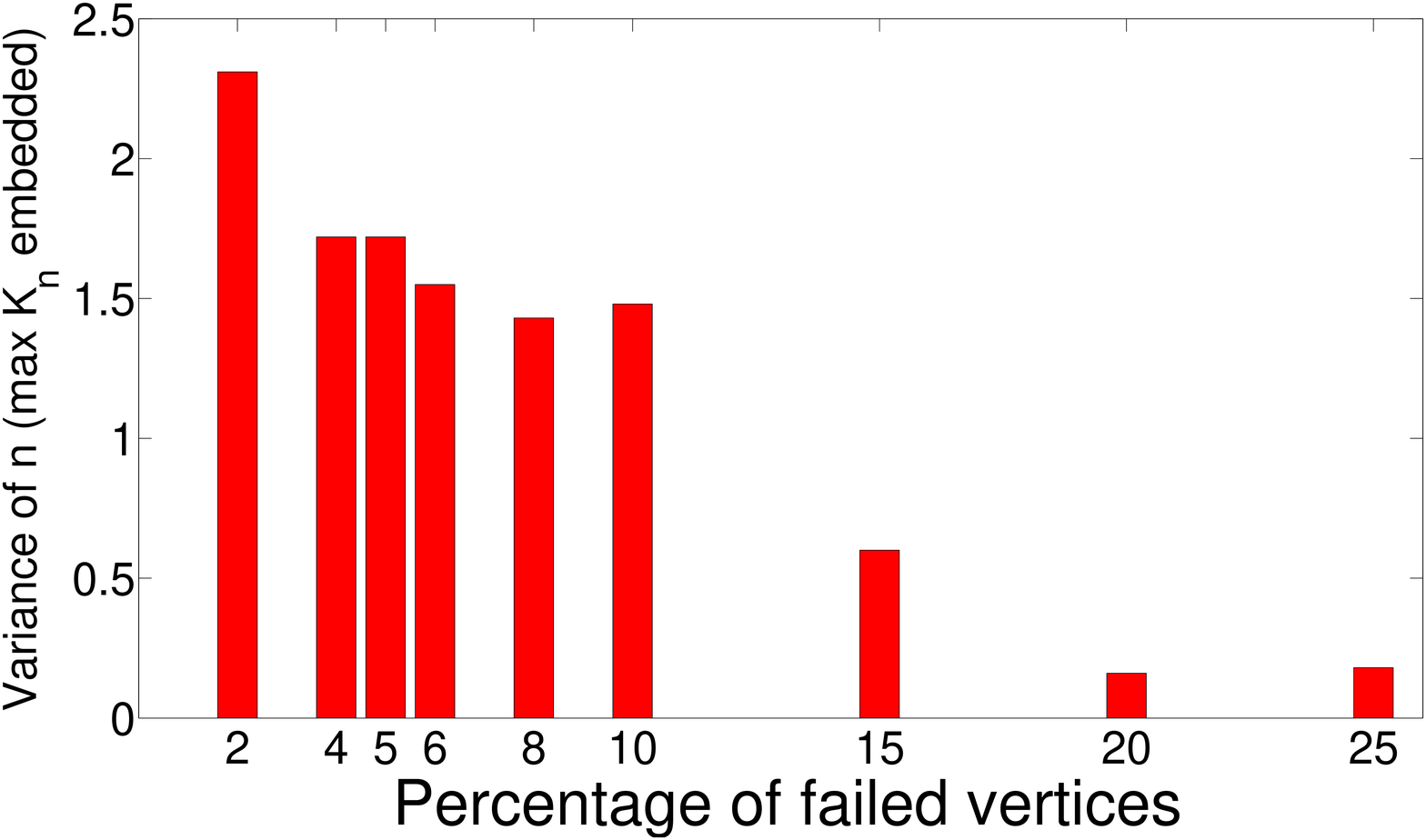}
}
\hspace{-0.1in}
\subfigure[Greedy algorithm]{
	\includegraphics[width=0.24\textwidth,height=.15\textwidth]{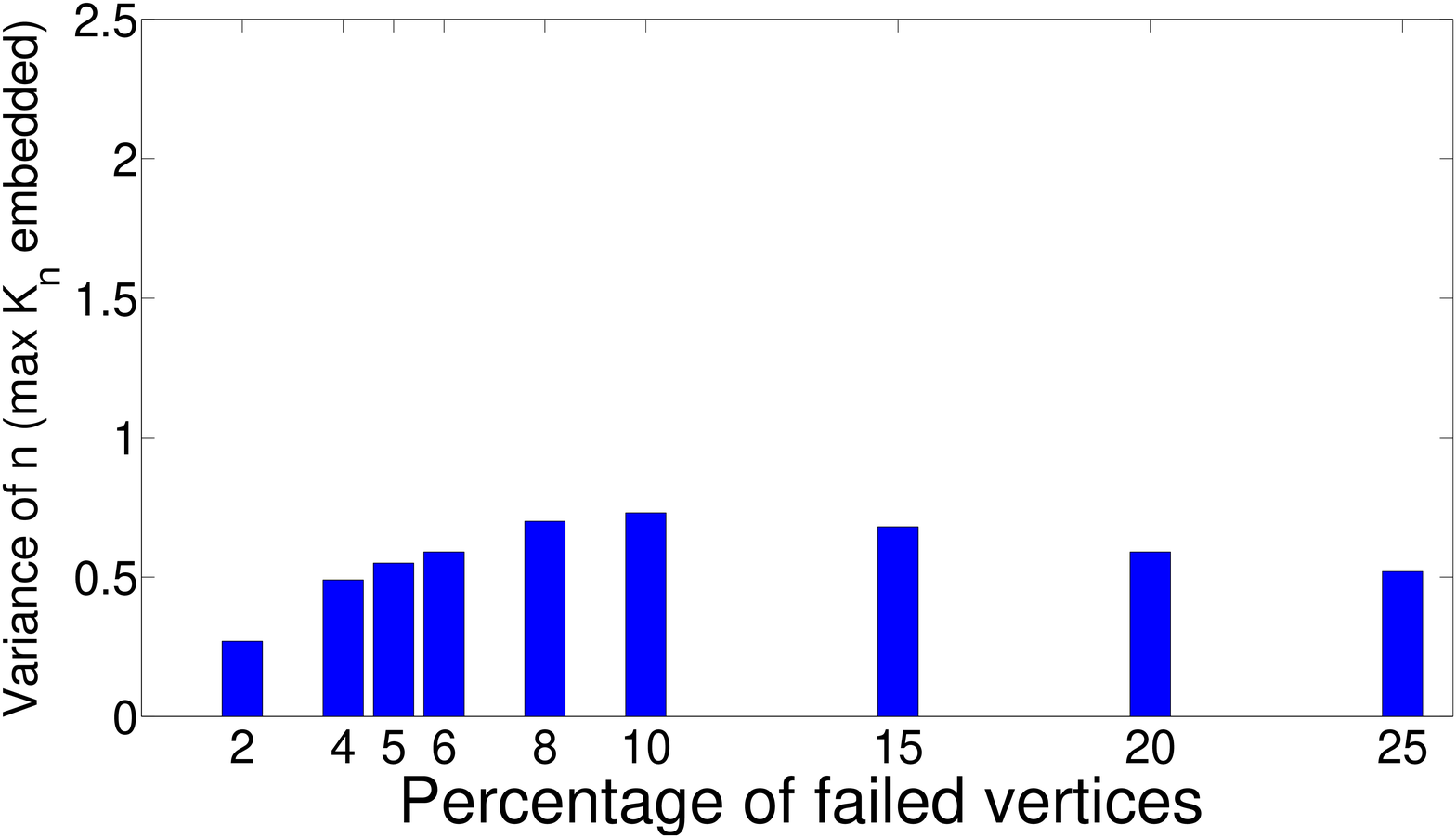}
}
\caption{Variances of fault-tolerant embedding algorithms (with flip and drop-down) on the $4\times4$ grid.}
\label{fig:variancehist}
\end{figure}

\begin{figure}[hb!]
\centering
\subfigure[Fallback embedding algorithm]{
	\includegraphics[width=.459\textwidth]{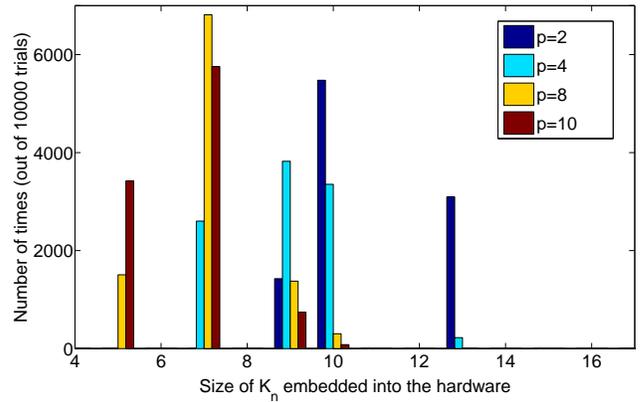}
}
\subfigure[Greedy embedding algorithm]{
	\includegraphics[width=0.459\textwidth]{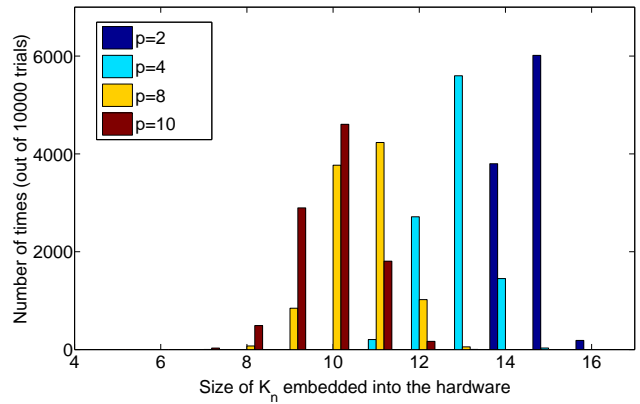}
}
\caption{Histograms for 10,000 trials of the fallback (left) and greedy (right) embeddings with flipping and drop-down on the $4\times4$ grid at $p=2,4,8,$ and 10 percent failure of the nodes.}
\label{fig:embedhist}
\end{figure}

\par
We have also analyzed the variances in embeddability from these experiments. In the case of a single attempt at embedding, the distribution of embeddable graphs tends to be narrower than when using the drop-down scheme. For larger grid sizes and for higher percentages of failure, the variance of the single attempt falls to zero.  This is caused by the fact that the algorithm never does better than embedding a $K_5$ into a single, complete cell.  However, this happens less often for the drop-down embedding scheme, yielding  larger average $K_n$ with higher variances.  An example of this behavior is shown in Fig.~\ref{fig:variancehist} for the case of $F(4,4)$ when varying the percent failure rate. It is notable that while the variance of the fallback method is relatively large for small error rates, the greedy algorithm maintains a near constant, much lower variance across all failure rates.  In Fig.~\ref{fig:embedhist}, the distribution of achieved embeddings over 10,000 trials using the flip and drop down scheme on $F(4,4)$, with the percentage of failed vertices at $p=2,4,8,$ and 10, is shown.  The embeddings achieved by the greedy algorithm are both more clustered and larger than those achieved by the fallback algorithm.  With the added evidence of panel (b) in Figs.~\ref{fig:singlevdropdown} and \ref{fig:fallbackvgreedy}, this demonstrates the greedy approach is more robust in the presence of hard faults.

\section{Conclusions}
We have presented methods for adiabatic quantum programming that embed problem specific information into an underlying quantum logical fabric. Our methods include an embeddability analysis based on the treewidth of an $m$-by-$m$ lattice of $K_{c,c}$ unit cells, which is a generalization of existing adiabatic quantum hardware. This has provided bounds on the graphs that can be embedded in a predefined logical fabric and should be useful for guiding adiabatic quantum programmed implementations.
\par
In addition, we have presented two new methods for finding an embedding of a complete graph in faulty fabric. The first method handles failures by falling back to a set of smaller available unit cells, while the second searches for embeddings that minimize the number of affected logical qubits using matching within cells on the diagonals. The latter was shown to have greater power for programming implementations of arbitrary QUBO instances. Numerical studies of embeddability run against randomized failures further showed the relative robustness of the second algorithm and the remarkably smaller variance in embeddable graphs.
\par
In our study of embedding for adiabatic quantum programming, we have neglected any question regarding the subsequent computational complexity. The question of how a particular embedding algorithm impacts the complexity of the resulting AQO program is a point for future research. The current work, however, is expected to support uncovering the dependency of the computational complexity on both the embedding and parameter setting methods used. We believe that the embedding algorithms explored here, which provide a constructive approach to programming, will be useful for providing a consistent means of comparing the AQO algorithm across different problem sizes and hardware.

\section{Acknowledgments}
This work was supported by the Lockheed Martin Corporation under Contract No. NFE-11-03394. The authors thank Greg Tallant (Lockheed) for technical interchange and Daniel Pack (ORNL) for help preparing Figure 2. This manuscript has been authored by a contractor of the U.S. Government under Contract No.~DE-AC05-00OR22725. Accordingly, the U.S. Government retains a non-exclusive, royalty-free license to publish or reproduce the published form of this contribution, or allow others to do so, for U.S. Government purposes.
 
\bibliographystyle{IEEEtran}

\end{document}